\documentclass[useAMS,usenatbib]{mn2e} 
\usepackage{graphicx}
\usepackage{amsmath}

\title[SF quenching in dusty S0s in the SSC]{ACCESS IV: The quenching of star formation in a cluster population of dusty S0s}
\author[Haines et al.]{C. P. Haines$^{1,2}$, P. Merluzzi$^3$, G. Busarello$^3$, M. A. Dopita$^{4,5,6}$, G. P. Smith$^{2}$, \and F. La Barbera$^3$, A. Gargiulo$^7$, S. Raychaudhury$^{2}$, R. J. Smith$^{8}$ \\
$^{1}$Steward Observatory, University of Arizona, 933 North Cherry Avenue, Tucson, AZ 85721, USA; cphaines@as.arizona.edu\\
$^{2}$School of Physics and Astronomy, University of Birmingham, Edgbaston, Birmingham, B15 2TT, UK\\
$^{3}$INAF - Osservatorio Astronomico di Capodimonte, via Moiariello 16, I-80131 Napoli, Italy\\
$^{4}$Research School of Astronomy and Astrophysics, Australian National University, Cotter Road, Weston Creek, ACT 2611, Australia\\
$^{5}$Astronomy Department, King Abdulaziz University, P.O. Box 80203, Jeddah, Saudi Arabia\\
$^{6}$Institute for Astronomy, University of Hawaii, 2680 Woodlawn Drive, Honolulu, HI 96822, USA\\
$^{7}$INAF - Osservatorio Astronomico di Brera, via Brera 28, I-20121 Milano, Italy\\ 
$^{8}$Department of Physics, University of Durham, Durham DH1 3LE, UK\\
}

\begin{document}

\maketitle
\label{firstpage}
\date{Accepted 2011 July 13. Received 2011 July 12; in original form 2011 May 06}

\begin{abstract}
We present an analysis of the mid-infrared (MIR) colours of 165 70$\mu$m-detected galaxies in the Shapley supercluster core (SSC) at $z{=}0.048$ using panoramic {\em Spitzer}/MIPS 24 and 70$\mu$m imaging. While the bulk of galaxies show $f_{70}/f_{24}$ colours typical of local star-forming galaxies, we identify a significant sub-population of 23 70$\mu$m-excess galaxies, whose MIR colours ($f_{70}/f_{24}{>}25$) are much redder and cannot be reproduced by any of the standard model infrared spectral energy distributions (SEDs). These galaxies are found to be strongly concentrated towards the cores of the five clusters that make up the SSC, and also appear rare among local field galaxies, confirming them as a cluster-specific phenomenon. Their optical spectra and lack of significant ultraviolet emission imply little or no ongoing star formation, while fits to their panchromatic SEDs require the far-infrared emission to come mostly from a diffuse dust component heated by the general interstellar radiation field rather than ongoing star formation. Most of these 70$\mu$m-excess galaxies are identified as ${\sim}L^{*}$ S0s with smooth profiles.

We find that almost every cluster galaxy in the process of star-formation quenching is already either an S0 or Sa, while we find no passive galaxies of class Sb or later. Hence the formation of passive early-type galaxies in cluster cores must involve the {\em prior} morphological transformation of late-type spirals into Sa/S0s, perhaps via pre-processing or the impact of cluster tidal fields, before a {\em subsequent} quenching of star formation once the lenticular encounters the dense environment of the cluster core. In the cases of many cluster S0s, this phase of star-formation quenching is characterised by an excess of 70$\mu$m emission, indicating that the cold dust content is declining at a {\em slower} rate than star formation. We suggest that the excess 70$\mu$m emission during quenching is due to either: (i) a reduction of the star-formation efficiency as proposed within the morphological quenching scenario; or (ii) a 2--3$\times$ increase in the dust-to-gas ratio or metallicity of the remaining interstellar medium, as predicted by chemical evolutionary models of galaxies undergoing ram-pressure stripping or starvation. 

\end{abstract}


\section{Introduction}
\label{intro}

\setcounter{footnote}{3}

The evolution of a galaxy is a product of both nature and nurture -- both of its mass, and of the environment in which the galaxy finds itself. Understanding the relative importance of nature and nurture remains a key issue in extra-galactic astrophysics. These factors control the star formation rate (SFR) and the dynamical structure or morphology of the galaxy. In isolated galaxies, these observational quantities are largely determined by feedback processes acting within the interstellar medium (ISM) of the galaxy adjoined to its assembly history via mergers. In the harsh cluster environment, star formation activity in infalling spiral galaxies can be both enhanced through collisions \citep{moss,bretherton} or quenched by various physical mechanisms such as ram-pressure stripping, harassment or starvation which transform them into the passive lenticulars which dominate the dense cluster cores \citep[for reviews see e.g.][]{treu,boselli,haines07}. These environmental mechanisms produce the well-known star formation (SF)--density \citep[e.g.][]{dressler85,poggianti99,balogh00,haines09} and morphology--density \citep{dressler80,dressler97,smith05} relations. However, the dominant evolutionary pathway(s) which generate these relationships remain uncertain. 

Moreover, it remains unclear whether these two relations are driven by the same process, by independent processes acting on diverse time-scales, or if one effectively drives the other. While some of the diminution in star-formation seen in cluster galaxies with respect to the field population could be attributed to the different morphological compositions of the two environments, even at fixed morphology, the star-formation rate of cluster galaxies is found to be lower than in the field \citep{balogh00}, indicating separate processes at least partially shape these relations. 

Alternatively, \citet{martig} suggest that it is the morphological transformation of spirals into S0s and the growth of the stellar bulge which drives the quenching of star-formation in lenticulars by making the gas disk more stable against collapse and fragmentation into star-forming clumps, without requiring gas consumption or removal. This interpretation is supported by the trend for the molecular gas consumption time-scale to increase systematically with galaxy concentration index ($R_{90}/R_{50}$), recently observed by \citet{saintonge}. This quenching itself is not related to the environment of the galaxy, and so cannot alone explain the SFR--density relation, instead requiring some additional environmental process(es) capable of producing the prior morphological transformation of the galaxies, such as the impact of cluster tidal fields \citep{byrd}, or pre-processing within infalling groups \citep{balogh04b}.

Infalling spirals in local clusters show evidence for the ongoing removal of gas via ram-pressure stripping \citep{crowl}, leading to a rapid termination of star-formation from the outside-in, and subsequent fading of the disk component. However, S0s are found to differ from normal spirals due to higher bulge luminosities rather than fainter disks \citep{christlein}, requiring bulge growth during S0 formation, disfavouring ram-pressure stripping or starvation models. Instead the increased scatters seen in cluster S0 and spiral Tully-Fisher relations support mechanisms that kinematically disturb infalling spirals, such as merging or harassment \citep{moran07}. Such tidal interactions are capable of channelling material to a central bulge sufficient to produce the higher central mass densities ($V_{rot}^{2}/Gr_{e}$) seen in cluster spirals \citep{moran07} and ultimately the central surface brightnesses and stellar phase densities found in S0s. 

A number of studies have attempted to distinguish among the various environmental processes by identifying and classifying various promising classes of transition galaxies, such as post-starburst and/or E+A galaxies \citep{dressler83,poggianti99,mercurio04,mahajan,mahajan11}, interaction-induced starbursts \citep{moss,fadda}, and passive/red spirals \citep{vandenbergh,bamford}, indicating that more than one mechanism is required. 

A key diagnostic to understand the relevance of these proposed transition galaxies is mid-infrared (MIR) data to account for the dust-obscured star formation. Many ``post-starburst'' galaxies identified by their deep H$\delta$ absorption lines and lack of O{\sc ii} emission were revealed by mid-infrared data to be dusty starbursts \citep{duc,dressler09}, whose age-dependant dust obscuration preferentially hides the young stars responsible for the O{\sc ii} emission above those older stars responsible for the Balmer absorption \citep{poggiantiwu}. These spectroscopic and MIR surveys revealed a ubiquitous population of starburst and post-starburst ${\sim}L^{*}$ spirals in $z{\sim}0.5$ clusters \citep{couch,geach,dressler09}, which are absent in present epoch clusters \citep{poggianti99}. They have been identified as the likely progenitors of present day cluster S0s, whose populations have risen dramatically over the last 5\,Gyr, empirically replacing the spirals \citep{dressler97,treu,smith05}. 
Additionally, interaction-induced starbursts are often highly obscured \citep{mihos}, while many red sequence galaxies (including spirals) have been revealed to be actively star-forming, their red colours being attributed to dust obscuration rather than being passively-evolving \citep{wolf05,wolf,haines10}. 

Most previous studies of environmental effects on galaxies have concentrated on their impact on morphology and star formation. However, given that the environmental processes behind the SF--density relation are likely to act {\em directly} on the ISM rather than the star forming regions themselves, to fully understand the origin of the SF-density relation requires also measures of these impacts of environment on the ISM which subsequently drive it. To date, a detailed knowledge of the impact of environment on the ISM and dust content, and how this drives the SFR--density relation is still largely lacking. 

The most accessible probe of the ISM in distant galaxies is via far-infrared observations of the thermal emission from their dust contents. Using early {\em IRAS} \citep{iras} and sub-mm observations \citet{desert} modelled the origin of the far-infrared emission in terms of two main components (ignoring PAHs here): a warm component of small dust grains in H{\sc ii} regions heated by star formation; and a cool cirrus component of large dust grains heated by the interstellar radiation field. \citet{lonsdale} estimated that this cool, cirrus component contributes as much as 50--70 per cent of the far-IR flux, while \citet{sauvage} revealed a strong morphological trend, with a decreasing cirrus contribution from ${\sim}85$ per cent for Sa galaxies to just 3 per cent for Sdm galaxies. 
The unprecedented wavelength coverage and sensitivity of {\em Herschel} \citep{pilbratt} allows us to extend these analyses to large samples of galaxies. \citet{kramer} demonstrate the requirement of a two-component dust model to fit the 24--500$\mu$m spectral energy distribution (SED) of M33, comprising a hot component (${\sim}6$0K) tracing star-formation and a dominant cold component of large dust grains heated by the general interstellar radiation field. This second component can in effect measure the fuel available for star formation, and a comparison of the two components allows us to measure the star-formation efficiency, or gas consumption time-scale. 
      
Using {\em IRAS}, \citet{doyon} found that H{\sc i}-deficient galaxies in the Virgo cluster had lower 60 and 100$\mu$m fluxes, and cooler far-infrared colours than those with normal H{\sc i} content. They interpreted this as evidence for the same two-component model for the far-infrared emission. In the cluster environment, the warm star-formation component is diminished in proportion to the H{\sc i} deficiency suggesting that star-formation is regulated by the amount of available gas, while the cool component from diffuse dust is reduced by a smaller factor as dust is stripped from the galaxy. {\em ISO} extended the far-infrared coverage beyond that of {\em IRAS} to 240$\mu$m, revealing the presence of a significant cold dust component (15--20K) in a complete sample of 38 late-type Virgo cluster galaxies \citep{popescu02}. 

While the mid-infrared coverage of {\em Spitzer}/MIPS is dominated by the reprocessed dust emission from current star formation, particularly at 24$\mu$m, for the 70$\mu$m band typically ${\sim}4$0\,per cent of the integrated light from galaxies can be attributed to dust heated by the diffuse interstellar radiation from evolved stars \citep{li10,bendo,calzetti10}. Hence the mid-infrared colours ($f_{70}/f_{24}$) can be considered as a sensitive probe of the interrelation between the ISM and star formation in galaxies.

In this paper we examine the mid-infrared colours of galaxies in the Shapley supercluster core (SSC) at $z{=}0.048$, the most massive and dynamically active structure in the local ($z{<}0.1$) Universe \citep{raychaudhury}. In this paper the SSC refers to a chain of three merging clusters, A3558, SC\,1327-312 and SC\,1329-313, along with the two associated Abell clusters on either side, A3562 and A3556. In {\S}~2 we present the panoramic {\em Spitzer} and ancillary data obtained within the ACCESS\footnote{{\em http://www.na.astro.it/ACCESS}} multi-wavelength survey used in this work. In {\S}~3 we present the mid-infrared colours ($f_{70}/f_{24}$) of galaxies in the SSC, identifying a subpopulation with much redder infrared colours ($f_{70}/f_{24}{\ga}25$) than can be reproduced by any of the standard infrared SED models, before examining in detail the nature of these objects. In {\S}~4 we discuss possible origins of these 70$\mu$m-excess galaxies, which appear to be dust-rich S0s yet with extremely low SFRs, before examining their consequences for the possible evolutionary pathways in the formation of cluster S0s.
When necessary we assume \mbox{$\Omega_{M}{=}0.3$}, \mbox{$\Omega_{\Lambda}{=}0.7$} and \mbox{H$_{0}{=}70\,$km\,s$^{-1}$Mpc$^{-1}$}, such that at the distance of the Shapley supercluster 1\,arcsec is equivalent to 0.96\,kpc.

\section{Data}
\label{sec:data}

The Shapley supercluster core was observed by {\em Spitzer}/MIPS in medium scan mode over 27--30 August 2008 within the Cycle 5 GO programme 50510 (PI: C.P. Haines). The observations consist of five contiguous mosaics providing homogeneous coverage of a ${\sim}3$\,deg$^{2}$ region at both 24 and 70$\mu$m, with effective exposure times per pixel of 84 and 42s in the two bands respectively. The data were reduced and analysed as described in \citet[hereafter Paper II]{paper2}, producing catalogues which are 90\% complete to 0.35\,mJy at 24$\mu$m \citep[corresponding to $L_{IR}{=}7.5{\times}10^{8}\,L_{\odot}$ or ${\rm SFR}{=}0.05\,{\rm M}_{\odot}{\rm yr}^{-1}$ based on the templates of][]{rieke09} and 25\,mJy at 70$\mu$m (corresponding to $L_{IR}{=}5.7{\times}10^{9}L_{\odot}$).
The MIR sources are then cross-correlated with optical $B$- and $R$-band imaging from the Shapley Optical Survey \citep[SOS:][]{sos1,sos2} and near-infrared $K$-band imaging \citep[Paper I]{merluzzi}, which cover the same region (see Fig.~1 in Paper II; Fig.~\ref{spatialircols}). 

We have spectroscopy of 814 supercluster members, of which 396 and 165 are detected at 24$\mu$m and 70$\mu$m respectively. Of the 814 galaxies with redshifts placing them in the SSC, 415 come from the AAOmega-based spectroscopic survey of \citet{smith07}, for which high signal-to-noise spectra (median 60/{\AA}) were obtained via 8\,hr long exposure times. The 580V grating on the blue arm of the spectrograph provided wavelength coverage over 3700--5800{\AA} sampled at 1{\AA} per pixel, while the 1000R grating on the red arm covered 5800--7300{\AA} at 0.6{\AA} per pixel (1.9{\AA} FWHM).

\begin{figure}
\centerline{\includegraphics[width=80mm]{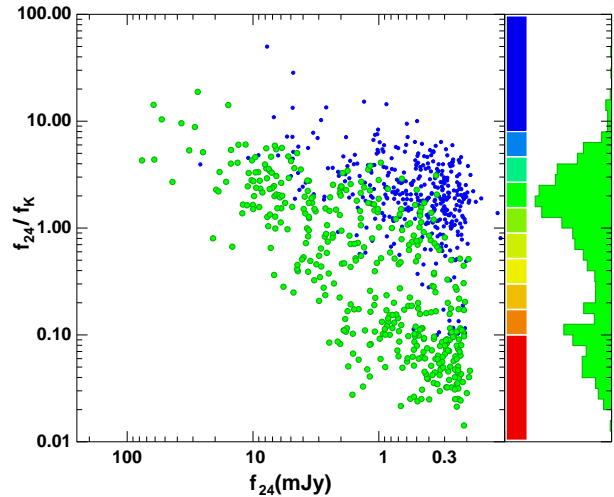}}
\caption{The infrared colour $f_{24}/f_{K}$ versus $f_{24}$ for Shapley supercluster galaxies identified spectroscopically (large green symbols) and photometrically according to their UV-optical colours (blue symbols).
The histogram shows the distribution of $f_{24}/f_{K}$ colours of the 24$\mu$m detected supercluster galaxies, demonstrating the clear bimodality in the infrared colours of supercluster galaxies. The vertical band indicates the colour scheme used in Fig.~\ref{mircols} to distinguish galaxies according to their  $f_{24}/f_{K}$ flux ratio.}
\label{optmir}
\end{figure}

In \citet[hereafter Paper III]{paper3} we found a clear bimodality in the infrared colour $f_{24}/f_{K}$ of SSC galaxies (shown here in Fig.~\ref{optmir}). 
This ratio is a proxy for the SFR per unit stellar mass, that can be used to separate passive ($f_{24}/f_{K}{\simeq}0.1$) and star-forming ($f_{24}/f_{K}{\simeq}2$) galaxies as discussed in detail in Paper III. This diagnostic also allows us to identify galaxies in between the two sequences, located in the  infrared analogue of the ``green valley'' ($f_{24}/f_{K}{\simeq}0.$15--1.0) population of UV--optical surveys. These galaxies are interpreted as being in the process of having their star formation quenched.  

\section{Results}
\label{sec:ircold}

\subsection{Mid-infrared colours}

\begin{figure}
\centerline{\includegraphics[width=80mm]{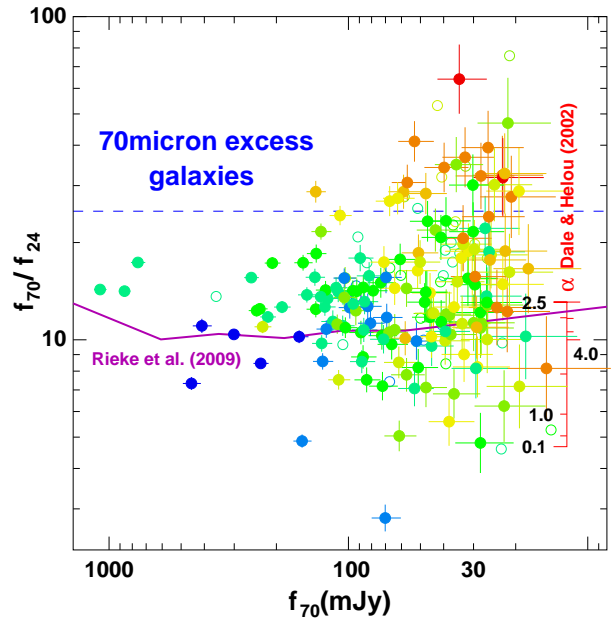}}
\caption{Mid-infrared colour $f_{70}/f_{24}$ versus 70$\mu$m flux of spectroscopically-confirmed (solid symbols) and photometrically selected (open symbols) Shapley supercluster galaxies. Each symbol is coloured according to its $f_{24}/f_{K}$ flux-ratio, as indicated by the coloured band of Fig.~\ref{optmir}. The magenta curve shows the predicted $f_{70}/f_{24}$ colour evolution of the luminosity-dependent infrared SEDs of \citet{rieke09}. The red scale shows the expected mid-infrared colours of the \citet{dh02} models for $\alpha{=}0$.1--4.}
\label{mircols}
\end{figure}

In Figure~\ref{mircols} we plot the mid-infrared colour $f_{70}/f_{24}$ versus 70$\mu$m flux of spectroscopically-confirmed (solid symbols) and photometrically-selected (open symbols) Shapley supercluster galaxies (see Paper II). Each symbol is coloured according to its $f_{24}/f_{K}$ flux-ratio, as indicated by the coloured band of Fig.~\ref{optmir}. Galaxies which belong to the star-forming sequence of Fig.~\ref{optmir} (green/blue symbols) generally have $f_{70}/f_{24}{\sim}$8--20, consistent with colours predicted by the luminosity-dependent SEDs of \citet{rieke09} (magenta curve) and also the \citet{dh02} models with $\alpha{\ga}1.5$. The latter are indicated by the red scale for $\alpha{=}0$.1--4, and show a turn over at $\alpha{\sim}2.5$. These SEDs are based on {\em IRAS} and {\em Infrared Space Observatory} observations of 69 normal star-forming galaxies \citep{dale} and further calibrated with sub-millimetre observations. This $\alpha$ value parametrises the distribution of dust mass as a function of heating intensity, as described in \citet{dh02}, in which values of $\alpha{\ga}2$ (as seen for most of our star-forming galaxies) indicate dust heating by low-intensity radiation typical of quiescent star formation extended over a spiral disk. The bluest mid-infrared colours ($f_{70}/f_{24}{\la}8$) imply smaller values ($\alpha{\la}1.5$), which correspond to more heating from stronger interstellar radiation fields, such as those produced by AGN or nuclear starbursts. Very few galaxies from our sample show such blue colours.

 We note here a general trend for galaxies to have cooler $f_{70}/f_{24}$ colours with decreasing $f_{24}/f_{K}$ flux-ratios. This would be expected for galaxies of a given mass/size as when their SFRs decline the intensitiy of the star-formation and ionizing UV radiation reduces, leading to a decline in the overall dust temperatures.

\begin{figure*}
\centerline{\includegraphics[width=170mm]{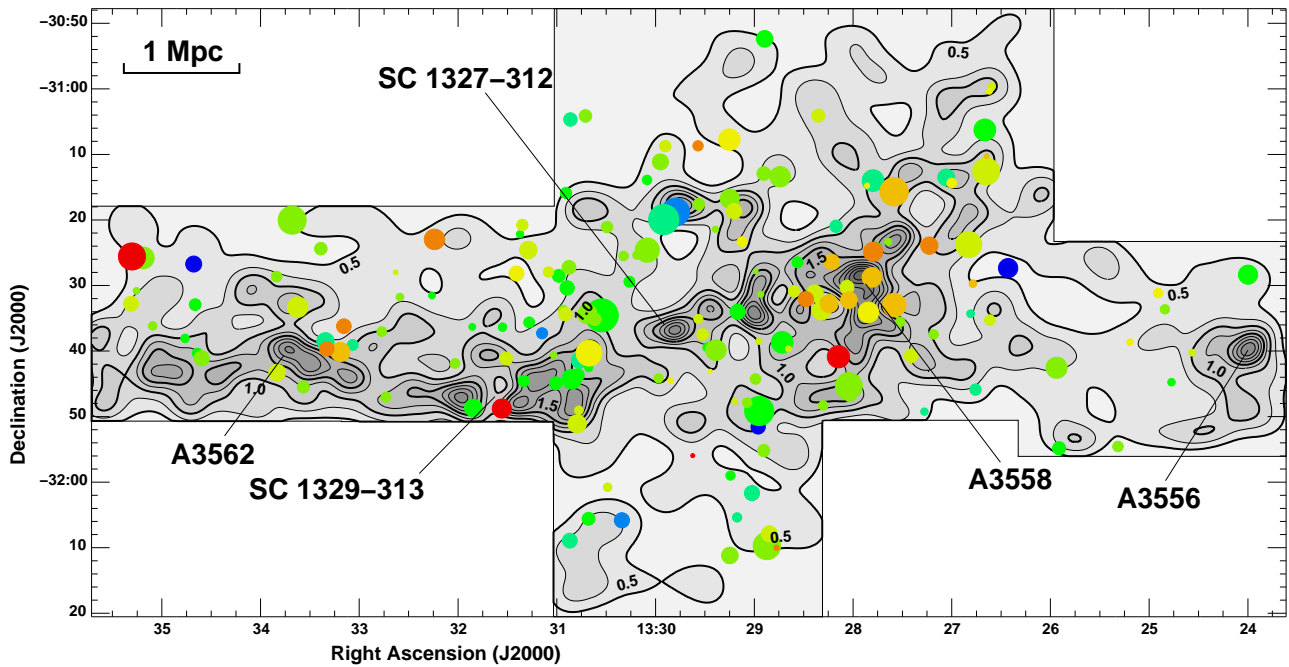}}
\caption{Spatial distribution of spectroscopically-confirmed 70$\mu$m sources across the Shapley supercluster. The symbols are coloured according to their $f_{70}/f_{24}$ mid-infrared colour from blue (${\la}7$) to orange/red (${\ga}30$), while the sizes scale with the $K$-band luminosity. The shaded region shows the coverage of WFI $B,R$-band imaging from the Shapley Optical Survey \citep{sos1}, from which we determine the morphologies. The grayscale contours indicate the surface density of $R{<}21$ galaxies \citep[Fig. 1 of][]{sos2}.}
\label{spatialircols}
\end{figure*}

We draw the reader's attention to the significant population of cluster galaxies with much redder infrared colours ($f_{70}/f_{24}{\ga}25$; above blue dashed line), that appear inconsistent with any of the \citet{rieke09} or \citet{dh02} SEDs at ${\ga}3{\sigma}$ significance levels. In total we identify 23 ``70$\mu$m-excess'' galaxies in the SSC having $f_{70}/f_{24}{>}25$, or $14{\pm}3$ per cent of the 70$\mu$m-detected SSC galaxies. 
We have visually checked the {\em Spitzer} maps and confirm that for each of these 23 sources corresponds to a clear, unambiguous detection in both 24 and 70$\mu$m images, with no possible issues with deblending or merging with other sources.  The spatial offsets between the 70$\mu$m and 24$\mu$m/$K$-band source positions of the 70$\mu$m-excess sources (${\sim}$1--3$^{\prime\prime}$) are entirely consistent with those found for the remainder of the 70$\mu$m sources of similar flux levels, and expectations of positional uncertainties based on their signal-to-noise levels. Our 70$\mu$m images are far too shallow to be affected by confusion noise \citep[$\sigma_{c}{\sim}0.3$mJy;][]{frayer}.  
As such we would expect that their anomalously red colours are not simply the result of photometric uncertainties, whereby the 70$\mu$m-excess is produced by low-S/N sources whose 70$\mu$m fluxes and $f_{70}/f_{24}$ flux ratios have been scattered up by statistical fluctuations.

To ensure that this is indeed the case, we created random samples based upon the observed 24$\mu$m fluxes and $f_{70}/f_{24}$ distribution of the supercluster galaxies, after censoring the 70$\mu$m-excess population. We then give each galaxy an artificial $f_{70}/f_{24}$ value assuming a log-normal distribution ($\mu{=}12.63$, $\sigma{=}0.144$\,dex) derived from the observed values, and then perturb both the 24$\mu$m and 70$\mu$m fluxes according to the observed photometric uncertainties (Figs~5 and~8 from Paper II). After taking into account the completeness of our observed catalogues as a function of actual 70$\mu$m flux, we estimate that we would expect ${\sim}3$ {\em detected} galaxies to have $f_{70}/f_{24}{>}25$ due to being in the tail of the intrinsic $f_{70}/f_{24}$ distribution, while ${\sim}3$ more would be scattered up into our sample by photometric uncertainties. We thus confirm that these 70$\mu$m-excess galaxies are a real subpopulation.

Of the 22/23 galaxies with $K$-band photometry, 16 can be characterized as being ${\sim}L^{*}$ galaxies, having $K$-band magnitudes in the range 11.3--12.8, placing them within ${\sim}1$\,mag of $M^{*}_{K}$ (Paper I), the remainder forming a faint-end tail. It is notable that most of these galaxies lie in the transition region between the passive and star-forming sequences of Fig.~\ref{optmir} (as indicated by their yellow/orange symbols in Fig.~\ref{mircols}), which could be interpreted as them being in the process of shutting down their star formation via their interaction with the dense cluster environment. Such galaxies appear similar to the population of cluster galaxies with unexpectedly high $f_{100}/f_{24}{>}30$ flux ratios identified in recent {\em Herschel} observations of clusters at $z{\sim}0.2$ \citep{rawle,pereira,smith10}. Overall, in terms of a mass-limited cluster galaxy population the 70$\mu$m-excess galaxies make up $8{\pm}2$ per cent of all $K{<}13$ ($M_{K}{<}M^{*}{+}1.3$) SSC galaxies. It is difficult to assess if and how this population extends to lower masses, as they are likely to drop below our 70$\mu$m survey limit beyond $K{\ga}13.5$, based on their expected infrared colours. 

We present the photometric properties of these 23 supercluster galaxies with $f_{70}/f_{24}{>}25$ in Table~\ref{catalogue}, including their position, morphology and bulge-to-total ratio (as determined in {\S}~\ref{morph}), $K$-band magnitude, $f_{70}/f_{24}$ and $f_{24}/f_{K}$ flux ratios and ${\rm NUV}{-}R$ colour.

\begin{table*}
\begin{tabular}{cccccccccccc} \hline
ID     & RA(J2000)   & Dec(J2000) & Morph& (B/T) & Velocity & $K$ & $f_{70}$(mJy) & $f_{70}/f_{24}$ & $f_{24}/f_{K}$ & $NUV{-}R$ \\ \hline
5587   &13:33:19.807 &-31:39:40.440 &S0  &   --  & 14000 & 12.819 & 22.74 & 32.31 & 0.144 & ${>}6.38$ \\
6222   &13:33:11.698 &-31:40:09.195 &E-S0& 0.757 & 14371 & 12.200 & 47.48 & 28.35 & 0.193 & 5.729\\
13898  &13:35:18.317 &-31:25:34.070 &E-S0& --    & 14863 &  --    & 64.33 & 41.81 & 0.000 & 6.220\\
22004  &13:32:14.399 &-31:22:57.368 &S0  & 0.515 & 14797 & 12.048 & 22.64 & 31.78 & 0.072 & 5.513\\
30092  &13:33:09.527 &-31:36:12.070 &S0  & 0.631 & 14440 & 12.772 & 27.94 & 32.15 & 0.170 & 5.756\\
36655  &13:31:33.506 &-31:48:46.282 &E   & 0.796 & 13250 & 12.207 & 34.44 & 64.05 & 0.062 & 6.232\\
61446  &13:27:13.632 &-31:23:54.909 &E-S0& --    & 14246 & 12.418 & 39.83 & 34.14 & 0.165 & 5.524\\
63446  &13:27:47.799 &-31:24:50.303 &S0  & 0.982 & 14119 & 12.179 & 32.55 & 36.72 & 0.100 & 5.895\\
64370  &13:28:12.948 &-31:26:24.527 &Sa  & 0.156 & 14853 & 12.639 & 20.86 & 27.69 & 0.130 & 4.704\\
64703  &13:26:31.124 &-31:26:49.215 &S0  & --    & 13896 & 14.176 & 19.38 & 28.86 & 0.479 & 4.346\\
65317  &13:27:48.517 &-31:28:45.618 &S0  & 0.536 & 15709 & 12.141 & 58.85 & 28.72 & 0.224 & 5.847\\
66331  &13:26:47.152 &-31:29:43.419 &Sg  & 0.293 & 15250 & 14.196 & 24.55 & 30.19 & 0.590 &${>}4.96$\\
67258  &13:28:28.612 &-31:32:04.525 &S0  & 0.294 & 14199 & 12.718 & 26.08 & 39.32 & 0.123 & 5.496\\
67346  &13:27:34.943 &-31:32:59.260 &E-S0& 0.662 & 12815 & 11.748 & 56.74 & 30.67 & 0.141 & 5.281\\
67440  &13:28:14.518 &-31:32:47.027 &Sa  & --    & 14518 & 12.356 & 62.27 & 27.41 & 0.303 &${>}7.37$\\
67459  &13:28:02.561 &-31:32:13.551 &S0  & 0.699 & 13846 & 12.484 & 67.25 & 26.86 & 0.376 & 5.077\\
71694  &13:28:08.902 &-31:40:52.153 &S0  & 0.678 & 12233 & 11.874 & 52.95 & 41.10 & 0.110 & 6.112\\
108346 &13:29:15.186 &-31:07:44.365 &E   & 0.690 & 14554 & 11.955 &108.39 & 25.23 & 4.296 & 5.114 \\
109544 &13:29:34.225 &-31:08:40.936 &S0  & 0.619 & 14278 & 13.553 & 22.24 & 32.69 & 0.273 & 5.011\\
129757 &13:26:38.815 &-31:10:16.229 &I   & 0.035 & 14150 & 15.225 & 30.21 & 30.12 & 1.878 & 1.507\\
131628 &13:27:35.052 &-31:15:41.428 &S0-S& 0.330 & 15095 & 11.340 &136.96 & 28.72 & 0.249 & 5.821\\
143443 &13:29:37.430 &-31:55:57.080 &Sg  & 0.002 & 14238 & 15.523 & 21.56 & 46.85 & 1.134 & 2.272\\
149556 &13:28:46.500 &-32:10:00.286 &Sg  & 0.051 & 13710 & 14.970 & 35.44 & 34.85 & 1.506 & 2.068\\ \hline
\end{tabular}
\caption{Photometric properties of the supercluster members with $f_{70}/f_{24}{>}25$. Where no NUV detection is made, we adopt a lower limit based on the NUV completeness limit of 22.0.}
\label{catalogue}
\end{table*}

\subsection{The environments of 70$\mu$m-excess galaxies within the SSC}

We plot in Fig.~\ref{spatialircols} the spatial distribution of the 70$\mu$m-detected galaxies across the Shapley supercluster, colour-coded according to their $f_{70}/f_{24}$ flux-ratio from blue (${\la}7$) through the typical ratios of normal star-forming galaxies in green (10--20) and the 70$\mu$m-excess galaxies in yellow/orange/red ($f_{70}/f_{24}{\ga}25$). The 70$\mu$m-excess galaxies are much more clustered than the overall 70$\mu$m supercluster galaxy population, concentrated predominately towards the core of the richest cluster Abell 3558, along with others near the cores of both Abell 3562 and SC\,1329-313. 

\begin{figure}
\centerline{\includegraphics[width=70mm]{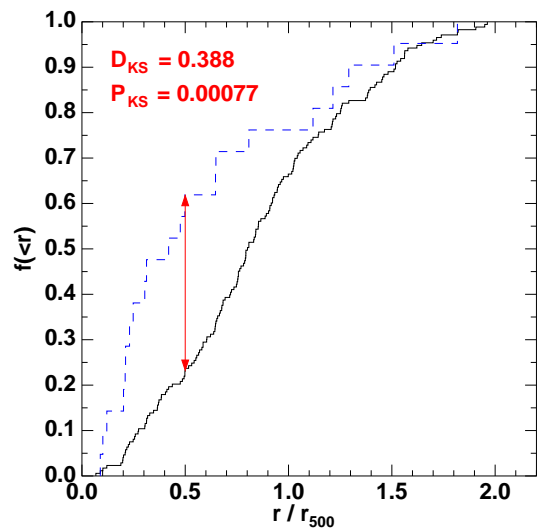}}
 \caption{Cumulative distribution of the distances of 70$\mu$m-excess (blue dashed curve) and all 70$\mu$m-detected (black solid curve) supercluster members from their nearest cluster in units of $r_{500}$.}
\label{ks}
\end{figure}

To quantify this apparent clustering towards the cluster cores, we plot in Fig.~\ref{ks} the cumulative distribution of distances of each 70$\mu$m-excess galaxy ($f_{70}/f_{24}{>}25$) from their nearest cluster in units of $r_{500}$ (blue dashed curve), where the $r_{500}$ values are estimated from mass density profiles derived from {\em Chandra} X-ray data \citep{haines09}. By comparing this to the corresponding distribution for all confirmed 70$\mu$m-detected supercluster members (black solid curve), we see that the 70$\mu$m-excess galaxies are on average at smaller cluster-centric radii, with 60\,per cent of them within $0.5\,r_{500}$ as opposed to 25\,per cent for the entire 70$\mu$m-detected population. Using the two-sample Kolmogorov-Smirnov test, we estimate the significance of this result, finding $P_{KS}{=}0.00077$ (3.36$\sigma$). Repeating this test, we find the 70$\mu$m-excess galaxies more concentrated towards the cluster cores also than the general supercluster population (i.e. all spectroscopic supercluster members), albeit at less significance ($P_{KS}{=}0.029$; 2.2$\sigma$).

\subsection{Comparison to field galaxy samples}

\begin{figure}
\centerline{\includegraphics[width=80mm]{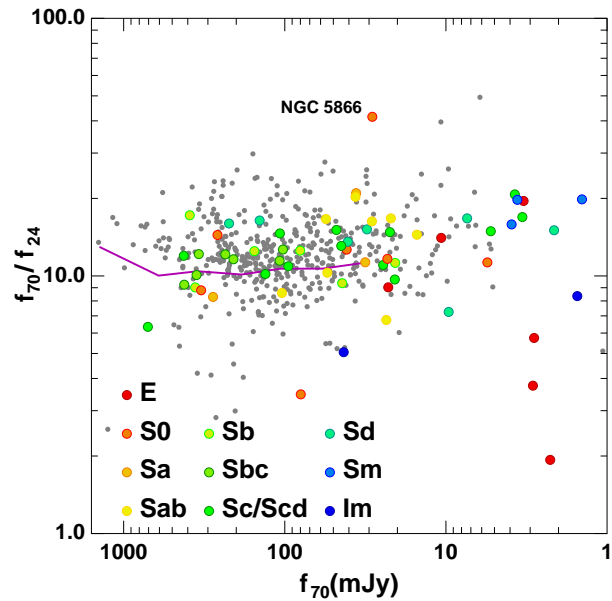}}
 \caption{Mid-infrared colour $f_{70}/f_{24}$ versus 70$\mu$m flux of local field galaxies from the SINGS and SWIRE surveys. The large coloured symbols identify galaxies from the SINGS sample, where the colour of the symbol indicates the morphological class from red (E) to blue (Im). The smaller grey symbols indicate galaxies from the SWIRE survey with spectroscopic redshifts ($z{<}0.12$) from SDSS. The $f_{70}$ flux is based upon the galaxy being shifted to the distance of the Shapley supercluster.}
\label{singsircols}
\end{figure}

Are these 70$\mu$m-excess galaxies purely a cluster phenomenon, or are they ubiquitous? To attempt to answer this, we have examined the infrared colours of local field galaxies taken from the SINGS \citep{sings} and {\em Spitzer} Wide-area InfraRed Extragalactic \citep[SWIRE;][]{swire} surveys. After correcting the infrared fluxes to the distance of the Shapley supercluster (to allow direct comparison with Fig.~\ref{mircols}), we plot in Fig.~\ref{singsircols} the $f_{70}/f_{24}$ infrared colour versus 70$\mu$m fluxes of the SINGS galaxies as large symbols, coloured according to their morphological class. The bulk of the SINGS galaxies have the same infrared colours as our observed star-forming sequence galaxies (green/blue symbols in Fig.~\ref{mircols}) with $8{\la}f_{70}/f_{24}{\la}20$. We identify just one galaxy, the S0 galaxy NGC\,5866, with the same anomalous red colour as our proposed 70$\mu$m-excess population. \citet{draine07} identified NGC\,5866 as a discrepant galaxy among their sample of 17 SINGS galaxies (mostly spirals) with SCUBA sub-mm photometry, due to its red $f_{70}/f_{24}$ colour, and in fitting a dust model to the infrared SED, required all of the dust to be exposed to a single diffuse interstellar radiation field, without any emission from dust heated by star-formation. \citet{li} do however find evidence for low-level star-formation (${\rm SFR}{\sim}0.05\,{\rm M}_{\odot}\,{\rm yr}^{-1}$) based on the detection of extended Paschen-$\alpha$ and H$\alpha$ emission, while {\em Spitzer}/IRS spectroscopy reveals clear PAH features at 7.7, 11.3 and 12.7$\mu$m due to star formation \citep{shapiro}.

\begin{figure*}
\centerline{\includegraphics[height=70mm]{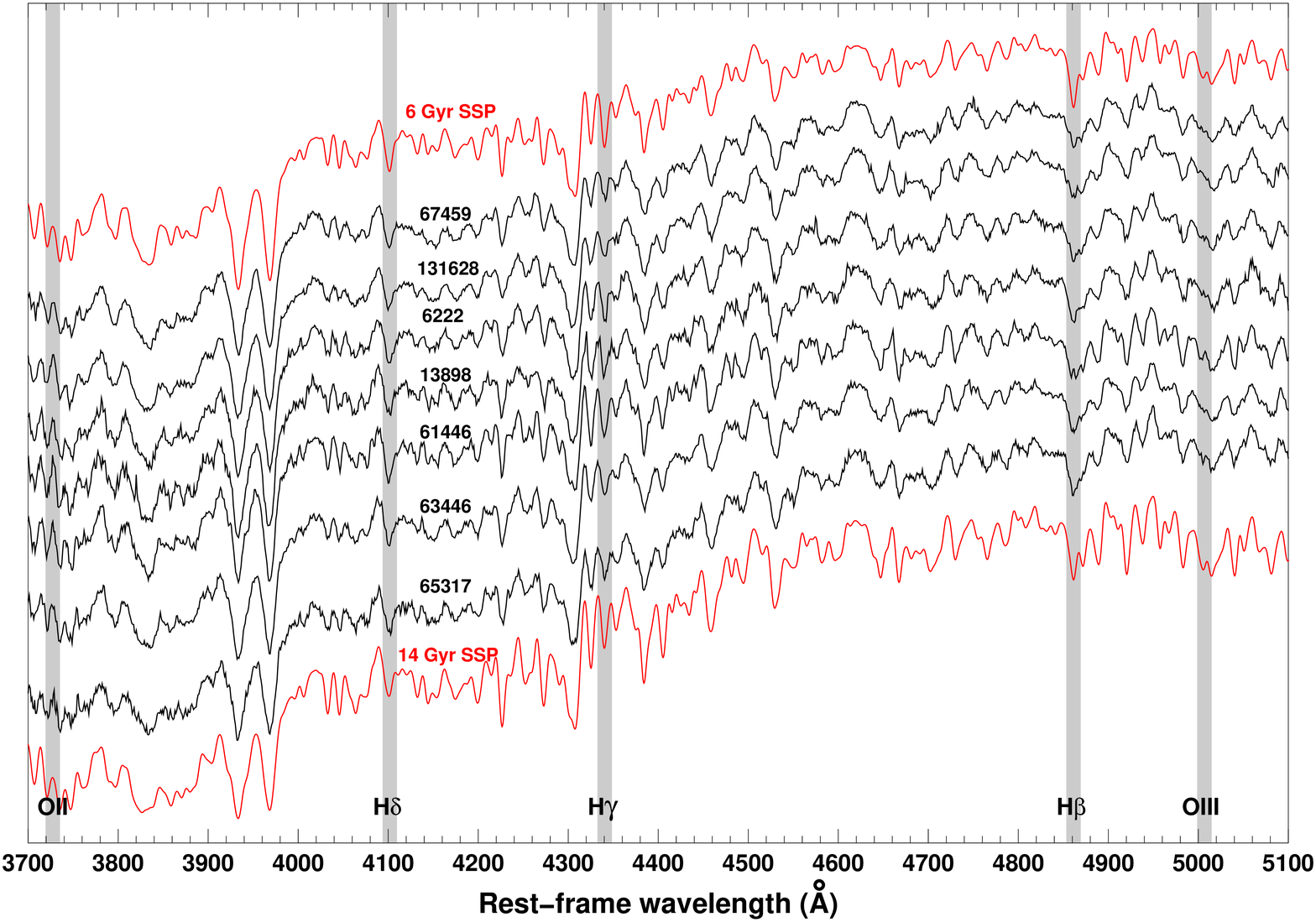}\includegraphics[height=70mm]{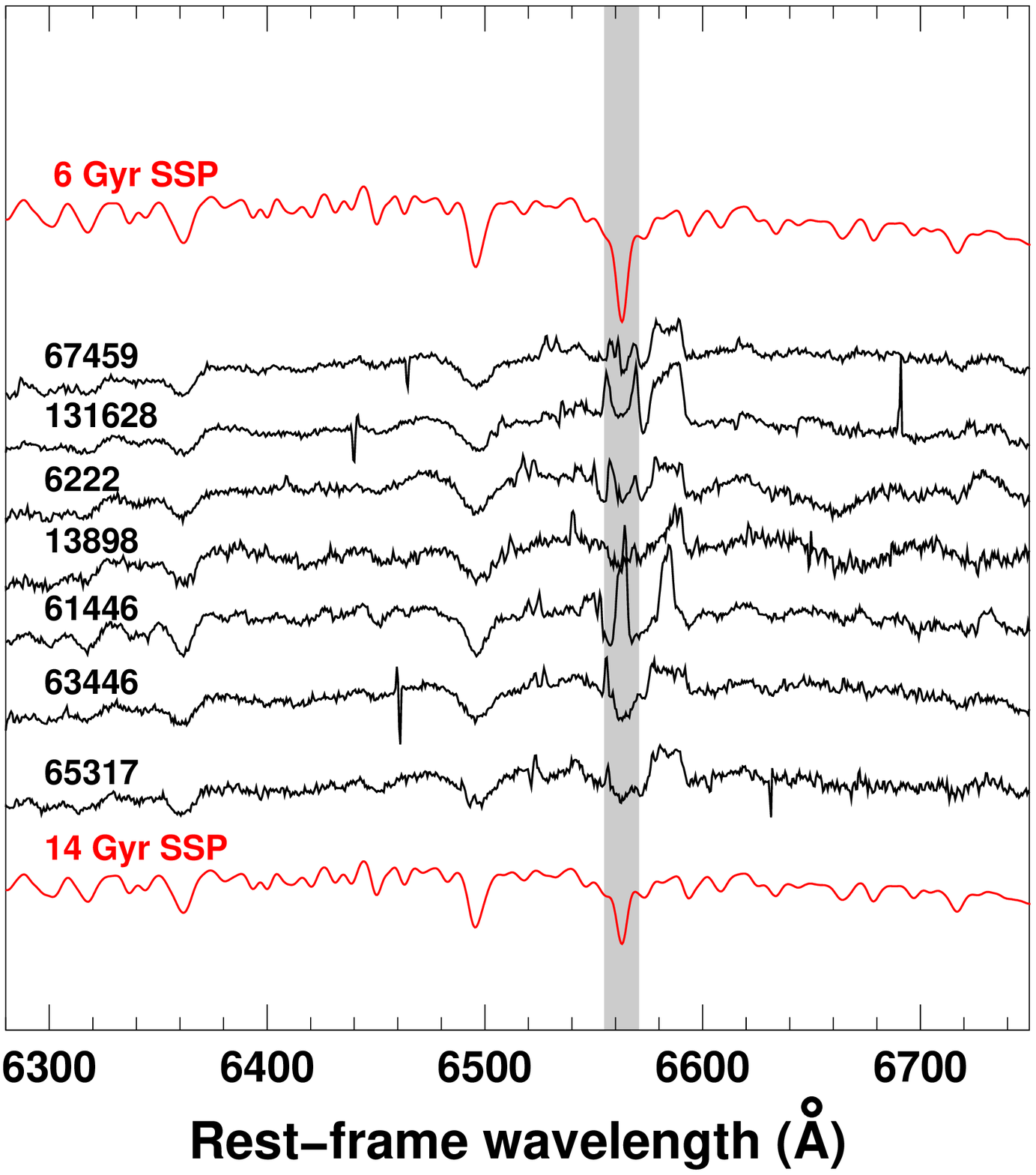}}
 \caption{Optical spectra (shifted to the rest-frame of the galaxy) of those 70$\mu$m-excess supercluster galaxies within the AAOmega sample of \citet{smith07} covering the wavelength range 3700--5100{\AA} ({\em left}) around the H$\alpha$ emission line ({\em right}). Also shown are two SSPs from the MILES stellar population models \citep{vazdekis} which bracket the properties of the observed spectra. The primary Balmer (H$\alpha$, H$\beta$, H$\gamma$, H$\delta$) and forbidden-line (O{\sc ii}, O{\sc iii}) indices are indicated by shaded vertical bars.}
\label{spectra}
\end{figure*}

We extend the field galaxy sample by matching the SWIRE 24$\mu$m and 70$\mu$m source catalogues for the Lockman and Elais N1/2 fields \citep{swire} to the SDSS DR7 optical spectroscopic and photometric catalogues \citep{sdssdr7}, identifying 455 $z{<}0.12$ galaxies with both 24$\mu$m and 70$\mu$m fluxes. Both the Lockman Hole and Elais N2 SWIRE fields fully lie within the SDSS DR7 footprint, covering 12.1 and 5.7\,deg$^{2}$ respectively, while about one-fifth of the 8.5\,deg$^{2}$ Elais N1 field also has SDSS spectroscopic coverage. We plot the FIR fluxes and colours of the SDSS-SWIRE matched sample as grey points in Figure~\ref{singsircols} after normalising the MIR fluxes to those which would be obtained if the source were at the distance of the Shapley supercluster. 
The obtained $f_{70}/f_{24}$ colours show a narrow range about the median value of 12.08, with $1\sigma$ and $2\sigma$ ranges of 9.25--16.5 and 5.3--24.8 respectively, fully consistent with the SINGS sample. We see a small population of hot IR-bright galaxies with $f_{70}{>}100$\,mJy and $f_{70}/f_{24}{<}7$, for which the bulk of the IR-emission is likely due to AGN. We confirm also the rarity of 70$\mu$m-excess FIR-bright galaxies in the field, with just five of 420--440 soureces ($1.2{\pm}0.5$ per cent) in the SDSS-SWIRE matched sample that would have been detected above the completeness limit of our Shapley 70$\mu$m imaging having $f_{70}/f_{24}{>}25$. This evident rarity of 70$\mu$m-excess sources in field surveys supports the hypothesis that these represent a cluster population.

\subsection{Spectra}

In Fig.~\ref{spectra} we present the spectra of all seven 70$\mu$m-excess galaxies from the AAOmega survey of \citet{smith07} having $f_{70}{>}30$mJy. 
In the left panel we present the spectra from the blue arm, covering the region from rest-frame O{\sc ii} to O{\sc iii}, including the 4000{\AA} break and the age-sensitive Balmer indices H${\delta}$, H$\gamma$ and H$\beta$ (marked by the shaded vertical bands). For comparison we show solar-metallicity simple stellar populations of ages 6 and 14\,Gyr from the MILES evolutionary synthetic models \citep{vazdekis}. Visually, these two models bracket and match well the overall continuum over this wavelength range as well as the primary age-sensitive features such as the 4000{\AA} break and the Balmer indices, suggesting that these galaxies are dominated by old stellar populations with ages in the range 6--14\,Gyr. This can be shown also quantitatively, as four out of these seven galaxies with EW(H$\alpha){<}0.$5{\AA} had mean stellar ages determined by \citet{smith07}, resulting in values of $6.3{\pm}0.5$\,Gyr (\#\,6222), $9.8{\pm}0.6$\,Gyr (\#\,13898), $11.0{\pm}0.5$\,Gyr (\#\,63446) and $7.7{\pm}0.7$\,Gyr (\#\,65317). There is also little if any evidence from the Rose Ca{\sc ii} indices \citep{leonardi} of ``frosting'' by small subpopulations of young stars (${<}1$\,Gyr), except perhaps for galaxy \#61446, while equally there is no evidence of post-starburst signatures in the form of deep Balmer absorption lines.

In the right panel we present the spectra from the red arm covering the region around H$\alpha$ for the same galaxies, again showing the 6 and 14\,Gyr old SSPs for comparison. Even low levels of ongoing star-formation should be detectable in the form of H$\alpha$ emission, but again we see little evidence for this in any of the galaxies except \#61446, although there is apparent filling in of the H$\alpha$ absorption expected of single old stellar populations. These galaxies have EW(H$\alpha$) in the range 0.3--3{\AA} and uncertainties of just 0.03--0.06{\AA} \citep{smith07}.

\begin{figure*}
\centerline{\includegraphics[width=88mm]{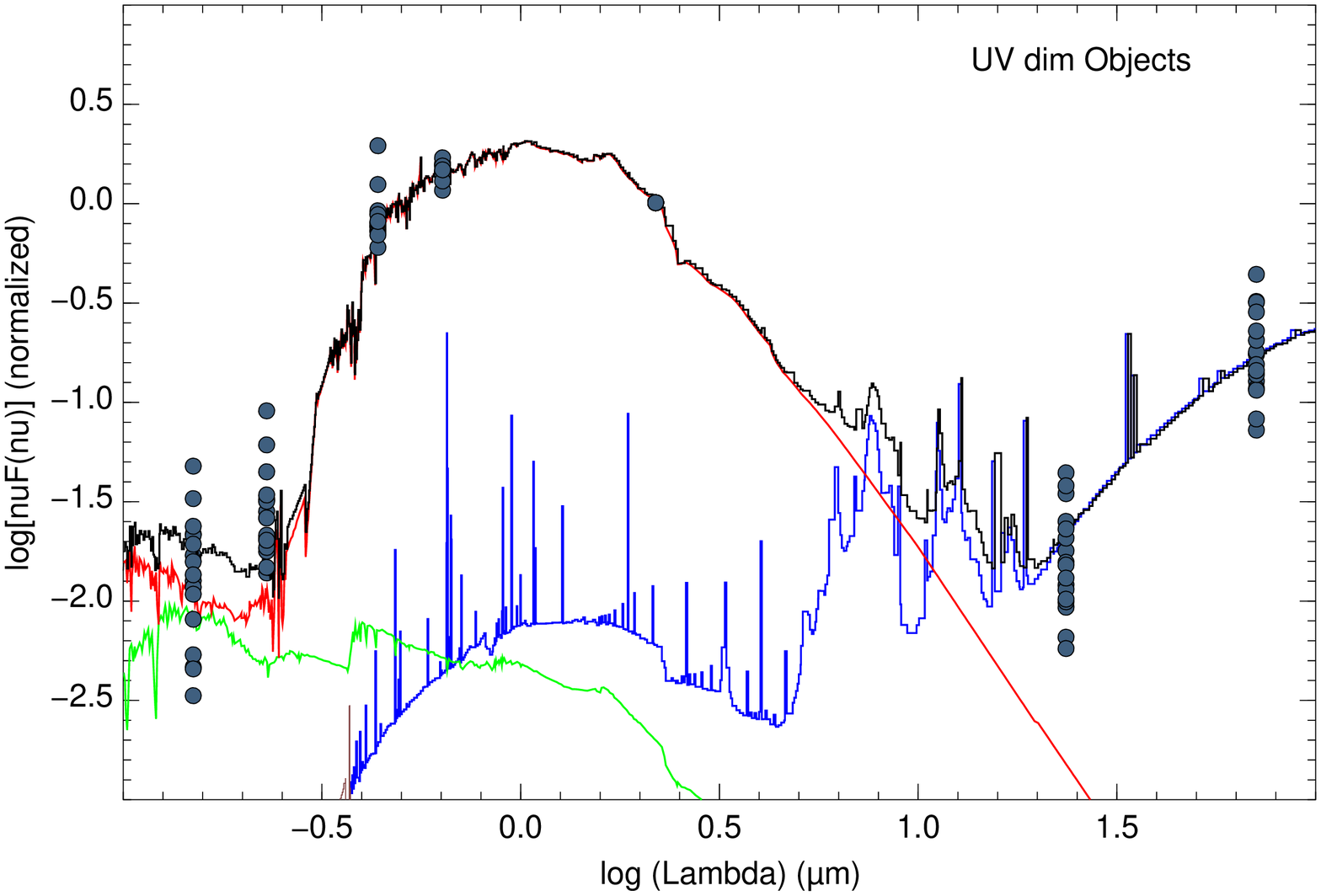}\includegraphics[width=88mm]{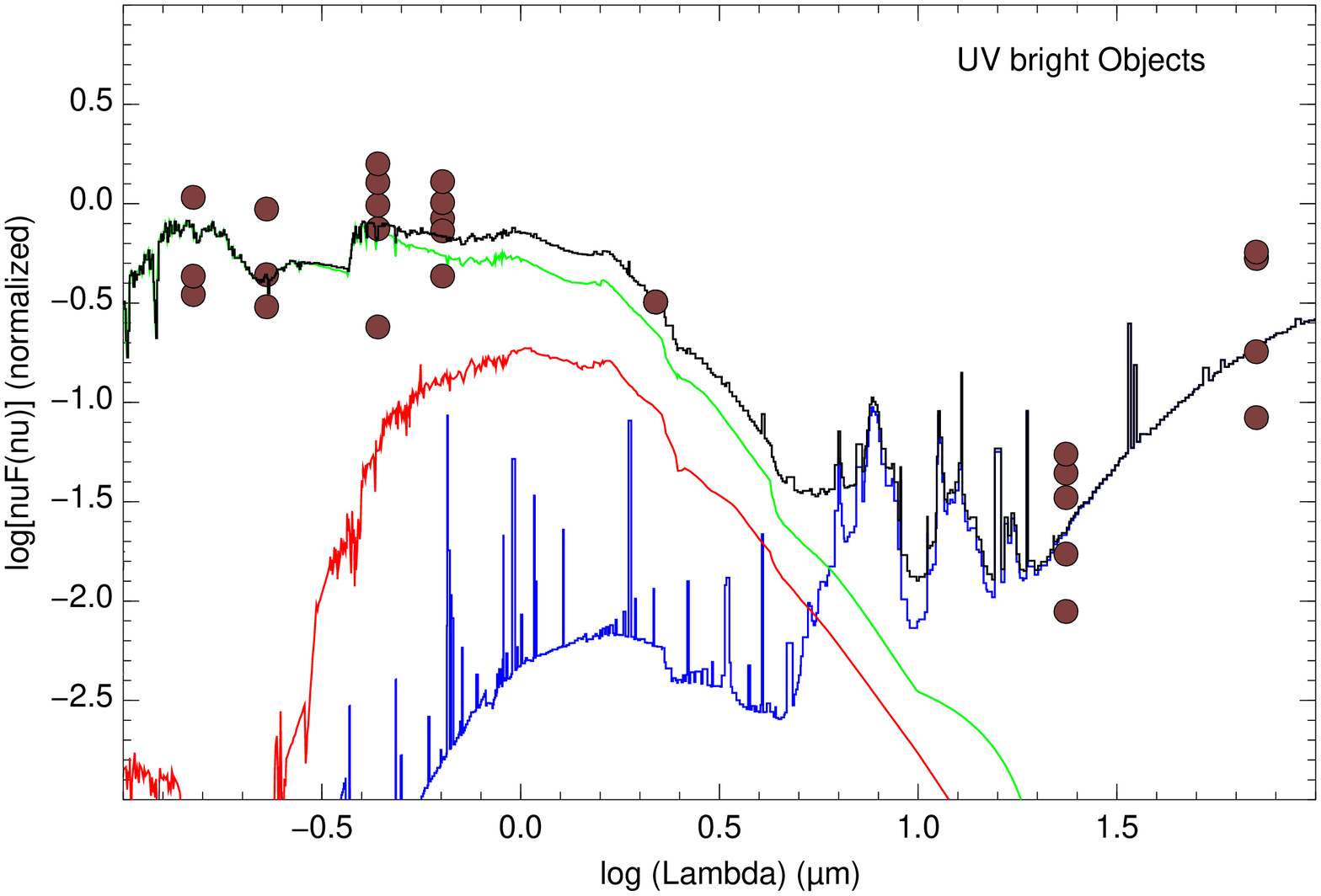}}
\caption{The best-fitting model (black curve) panchromatic FUV--FIR spectral energy distribution of the 70$\mu$m-excess galaxies normalized at 2.2$\mu$m (blue symbols) separated into those which are UV-faint ({\em left panel}) and UV-bright ({\em right panel}). The pan-spectral model SED is made up of three components: an 11\,Gyr old stellar population (red curve); young 10--100\,Myr old stellar population (green curve); and emission from the H{\sc ii} regions (blue curve).}
\label{sed}
\end{figure*}

One caveat to this is that the AAOmega 2\,arcsec diameter fibres cover just the central kpc of the galaxies, which can lead to severe aperture biases in the cases of galaxies with strong radial age or star formation gradients \citep{kewley}. In the case of early-type spirals or S0s this could imply sampling only the central bulge, missing extended star-formation occuring within the outer disk and spiral arms as demonstrated most clearly in \citet{haines08}. \citet{rawleifu} presented VIMOS integral field spectroscopy of three of the 70$\mu$m-excess galaxies (\#65317$\equiv$MGP\,1626; \#13898$\equiv$MGP\,3976, \#30092$\equiv$MGP\,3971). They found no evidence for young stellar populations in the outer regions of these galaxies. Indeed MGP\,1626 and MGP\,3971 have the two strongest {\em positive} age gradients in their sample of 19 galaxies, i.e. the central regions covered by the AAOmega fibres map the youngest stellar populations in the galaxies. These three galaxies show no statistical differences in the radial gradients of any of the stellar population parameters (age, metallicity, abundances) or velocity dispersion from the remainder of the sample (T. Rawle; private communication). This global lack of star formation in these galaxies is confirmed by their red ${\rm NUV}{-}R{\sim}$5--6 colours, placing them within the UV-optical red sequence \citep{haines08}. 

In summary, both spectroscopic (both fibre-based and integral field) and ultraviolet continuum observations of the 70$\mu$m-excess galaxies find them completely dominated by emission from evolved stars with little or no evidence of young stellar populations (${\la}1$\,Gyr).

\subsection{Panchromatic SEDs}

We now attempt to gain insight into the origin of the mid-infrared colours of these galaxies, by examining their panchromatic spectral energy distributions (SEDs). In Figure~\ref{sed} we plot the SEDs of all the 70$\mu$m-excess galaxies (with $K$-band photometry) from the far-ultraviolet to the far-infrared, after normalizing at 2.2$\mu$m (circular symbols). In order to efficiently model their SEDs, we have first split the population into two according to whether they appeared UV dim (${\rm NUV}{-}R{\ga}5$; left panel) or showed significant UV emission (${\rm NUV}{-}R{\sim}2$; right panel). In terms of numbers, the vast majority (19/22) of the 70$\mu$m-excess galaxies fall in the UV-dim sub-population, and so we will focus most attention on these.  

We have modelled the averaged panchromatic SEDs of the two sub-populations by the starburst models of \citet{groves}. These models combine the stellar spectral synthesis code {\sc Starburst}99 \citep{leitherer} and the photoionization code {\sc mappings iii} \citep{groves04}, which follows both the nebular emission and radiative transfer locally within individual H{\sc ii} regions. The expansion of these regions is driven by the mechanical energy input of their cluster stars and supernovae \citep{dopita05}. The FUV--FIR SEDs of the 70$\mu$m-excess galaxies galaxies were fit by models with three components: (i) an ensemble of young H{\sc ii} regions surrounding young clusters with ages ${<}1$0\,Myr enclosed by photodissociation regions and foreground attenuation by a dust screen (blue curve); (ii) a young stellar population with ages 10--100\,Myr modelling the young stars that have recently dispersed into the field from the active star-forming regions (green curve); and (iii) an old stellar population based on the 11\,Gyr template of \citet{bc03} to model the evolved stars unconnected with the current star formation episode (red curve). 

In the UV--optical we see little or no contribution from the current or recent star formation to the spectra for the vast majority of these galaxies ({\em left panel}), being essentially modelled by a simple stellar population of age 11\,Gyr. This is particularly so for 0.3--3\,$\mu$m, consistent with the observed AAOmega spectra being well described by 6--14\,Gyr old SSPs (Fig.~\ref{spectra}). Only in the FUV is any contribution from young (${<}100$\,Myr old) stars apparent, but even here it could be feasibly be excluded for many of the galaxies. This doesn't appear to be due to dust extinction, the underlying old stellar population is extinction free, while the modelled young 10--100\,Myr old component has $A_{V}{\sim}0.3$\,mag. The models predict a negligible contribution at UV-optical wavelengths from H{\sc ii} regions, with little or no detectable nebular emission. This is confirmed by the lack of significant H$\alpha$ emission in any of the galaxies in Fig.~\ref{spectra} or those covered by IFU spectroscopy \citep{rawleifu}. Moreover, the optical and nebular emission suggested by the blue curves is an artifact of the models, which connect implicitly the dust heating process to H{\sc ii} regions, producing nebular emission as a consequence of the ionizing UV radiation (heavily extincted) required to heat the dust sufficiently to produce the observed far-infrared emission. Given the clear lack of observed UV or nebular emission, the origin of the UV/optical radiation required to heat the dust component remains unclear.

The emission from H{\sc ii} regions is instead only detectable in the mid/far-infrared, demonstrating the importance of such data for fully understanding the processes which drive and quench star formation in galaxies. The H{\sc ii} component indeed is required to fit the observed emission in the {\em Spitzer}/MIPS 24$\mu$m and 70$\mu$m passbands, the predicted 24$\mu$m emission due to photospheric emission from the evolved stellar population (red curve) being an order of magntiude lower than that observed. The models also predict that emission from the H{\sc ii} regions should be detectable in the form of PAH emission at 6--20$\mu$m. Unfortunately, we do not have any photometry at these wavelengths, although we note again that NGC\,5866 from the SINGS sample does indeed show clear PAH features \citep{shapiro}.

One caveat is that {\em Spitzer} Infrared Spectrograph (IRS) observations of early-type galaxies have revealed diffuse, excess emission over 10--30$\mu$m, due to silicate emission from the dusty circumstellar envelopes of mass-losing asymptotic giant branch stars \citep{bressan}, which persists even for the very evolved stellar populations beyond 10\,Gyr old \citep{piovan}. While this component is not included in our models, and could in principle explain the 24$\mu$m fluxes, its emission should then decline rapidly beyond 30$\mu$m, and hence could in no way contribute significantly to the thermal emission detected at 70$\mu$m. 

We do note however that 3/23 galaxies from our sample do have significant UV emission from recent star formation ({\em right panel} of Fig.~\ref{sed}), having ${\rm NUV}{-}R{\sim}2$. These three also have $f_{24}/f_{K}{\sim}1.5$ ratios typical of normally star-forming galaxies, much lower $K$-band luminosities and later-type morphologies than the remainder of the sample. 
Given the rather diverse properties of these 3 galaxies with respect to all of the others, and the likelihood that many similar low-mass galaxies would be missed by our 70$\mu$m survey, it is difficult to assess the importance or relation of these galaxies with respect to the rest of the population. 

\begin{figure*}
\centerline{\includegraphics[width=34mm]{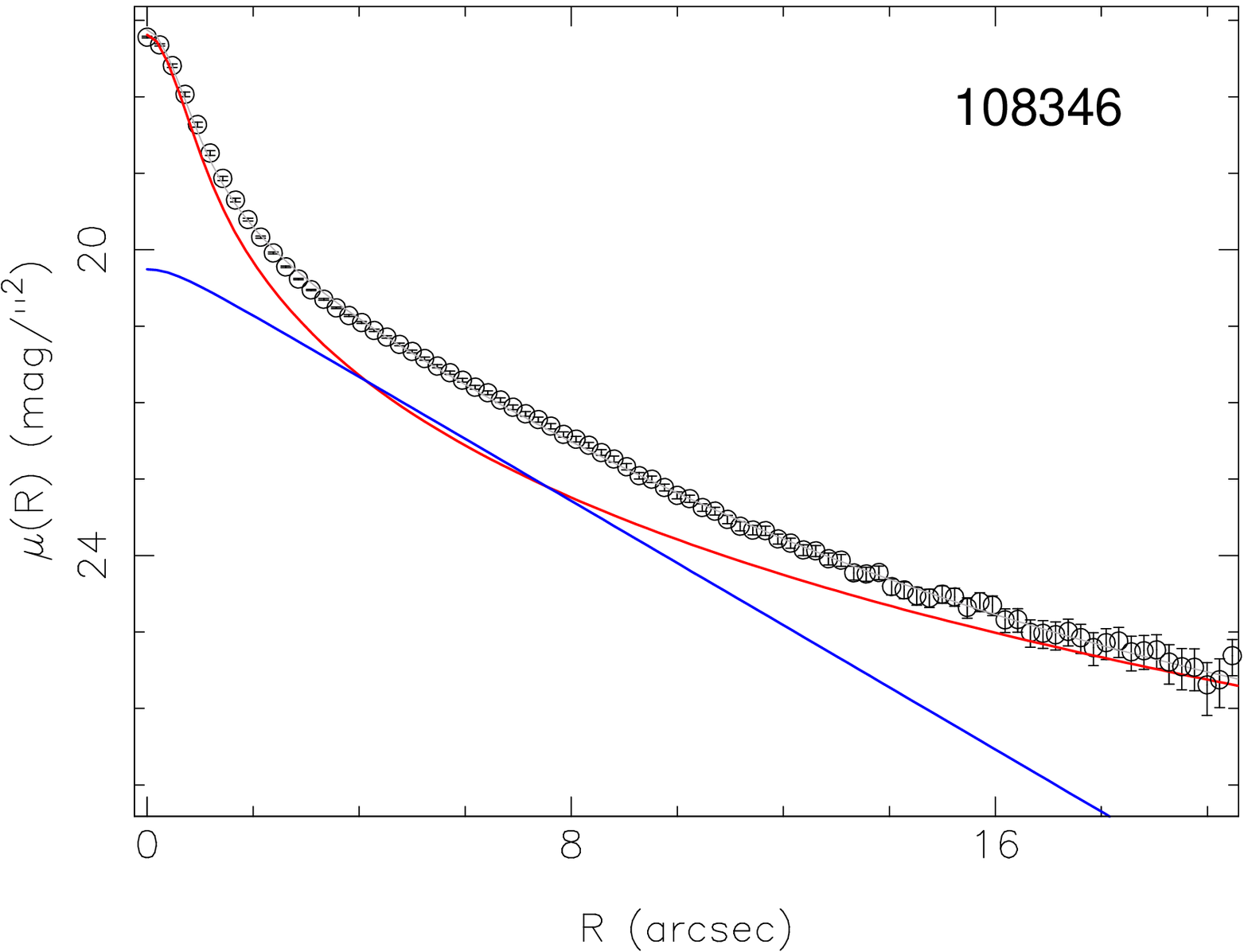}\includegraphics[width=34mm]{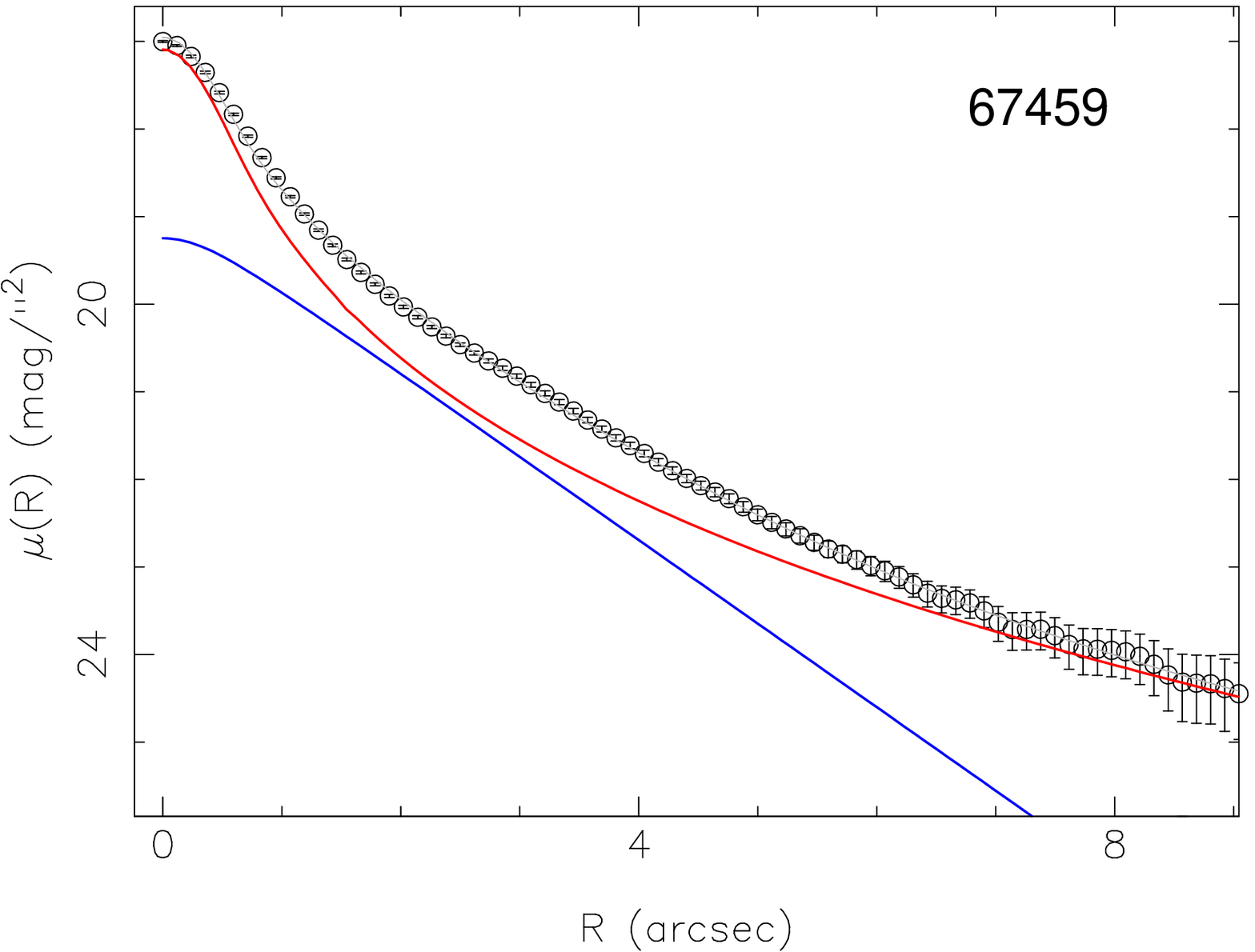}\includegraphics[width=34mm]{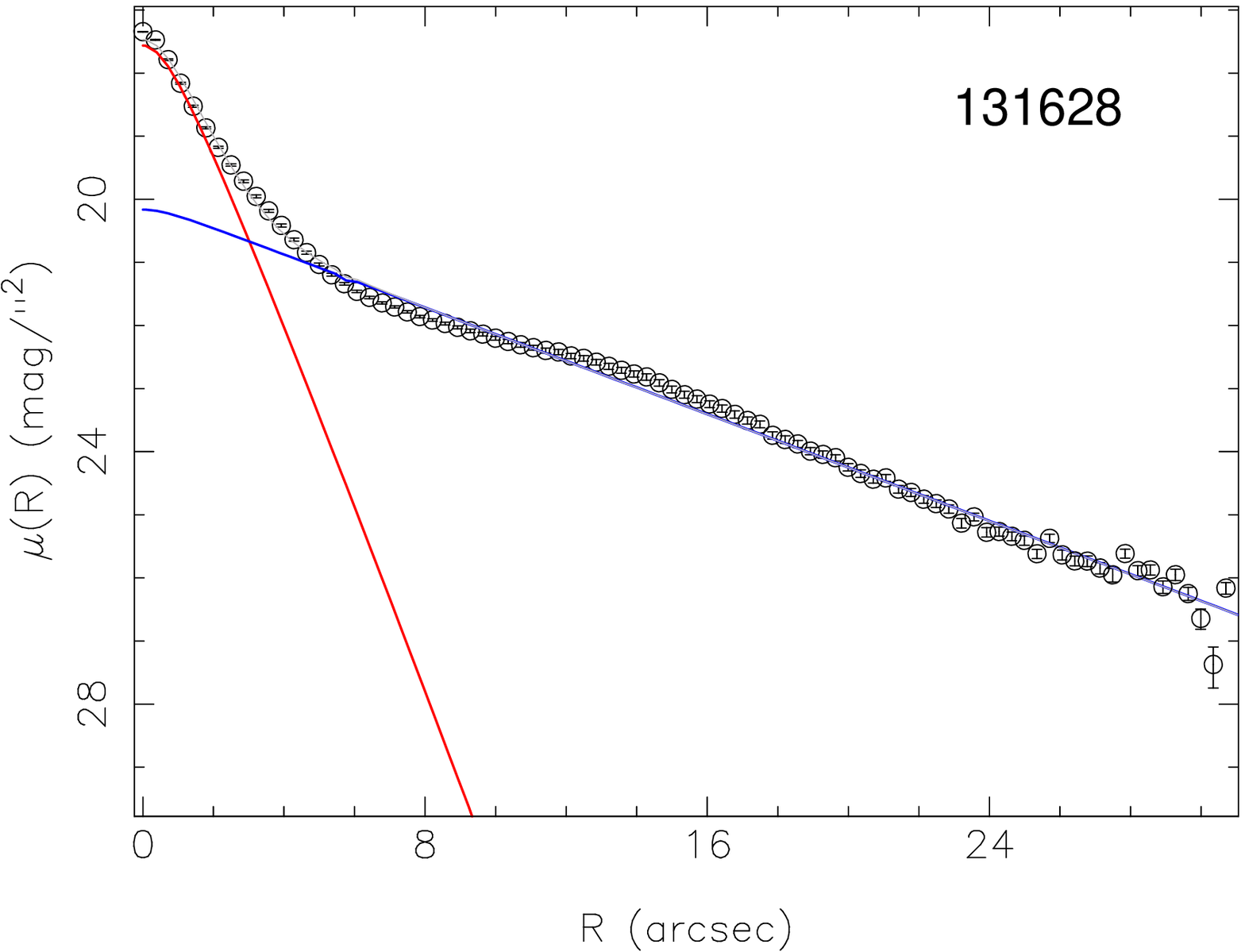}\includegraphics[width=34mm]{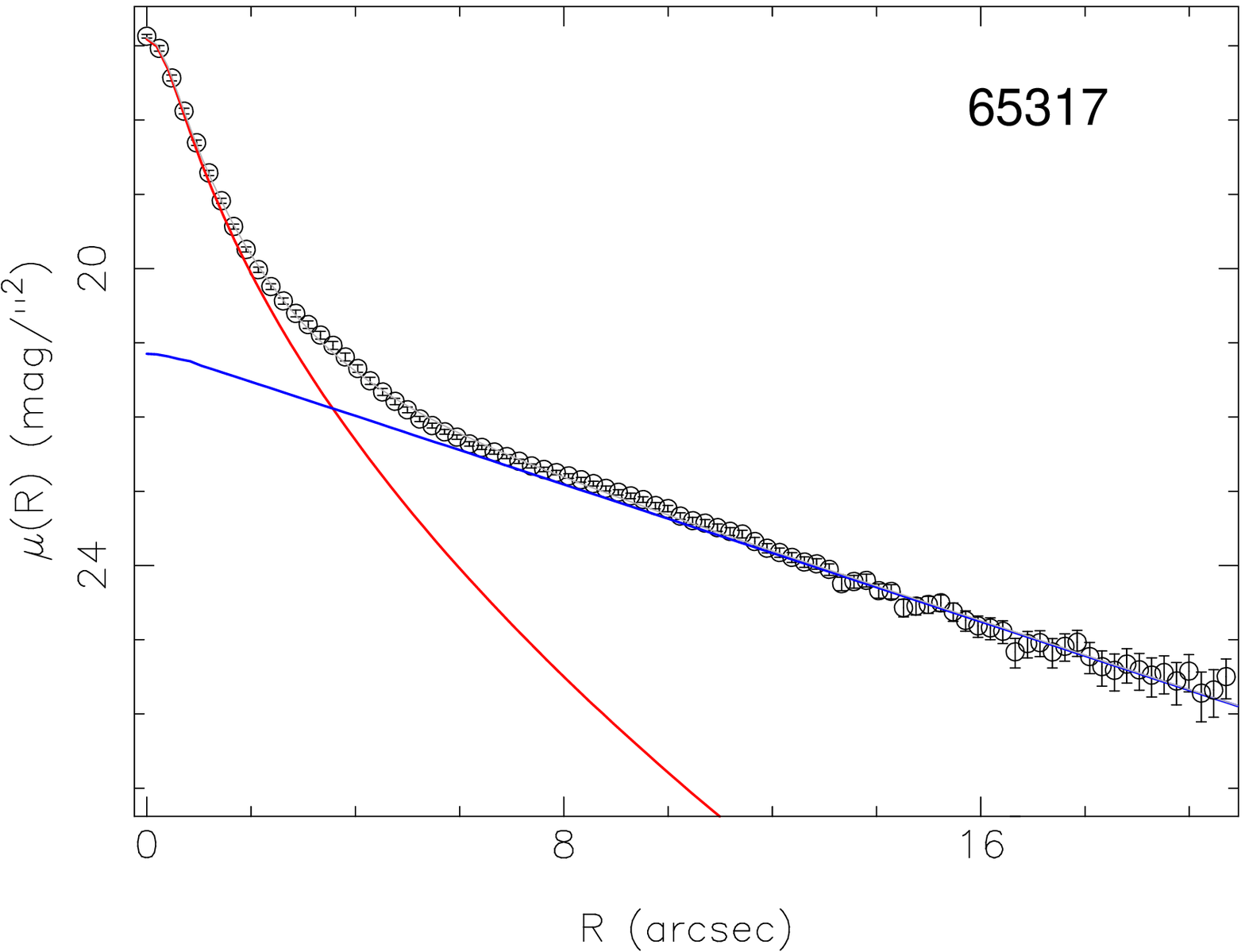}\includegraphics[width=34mm]{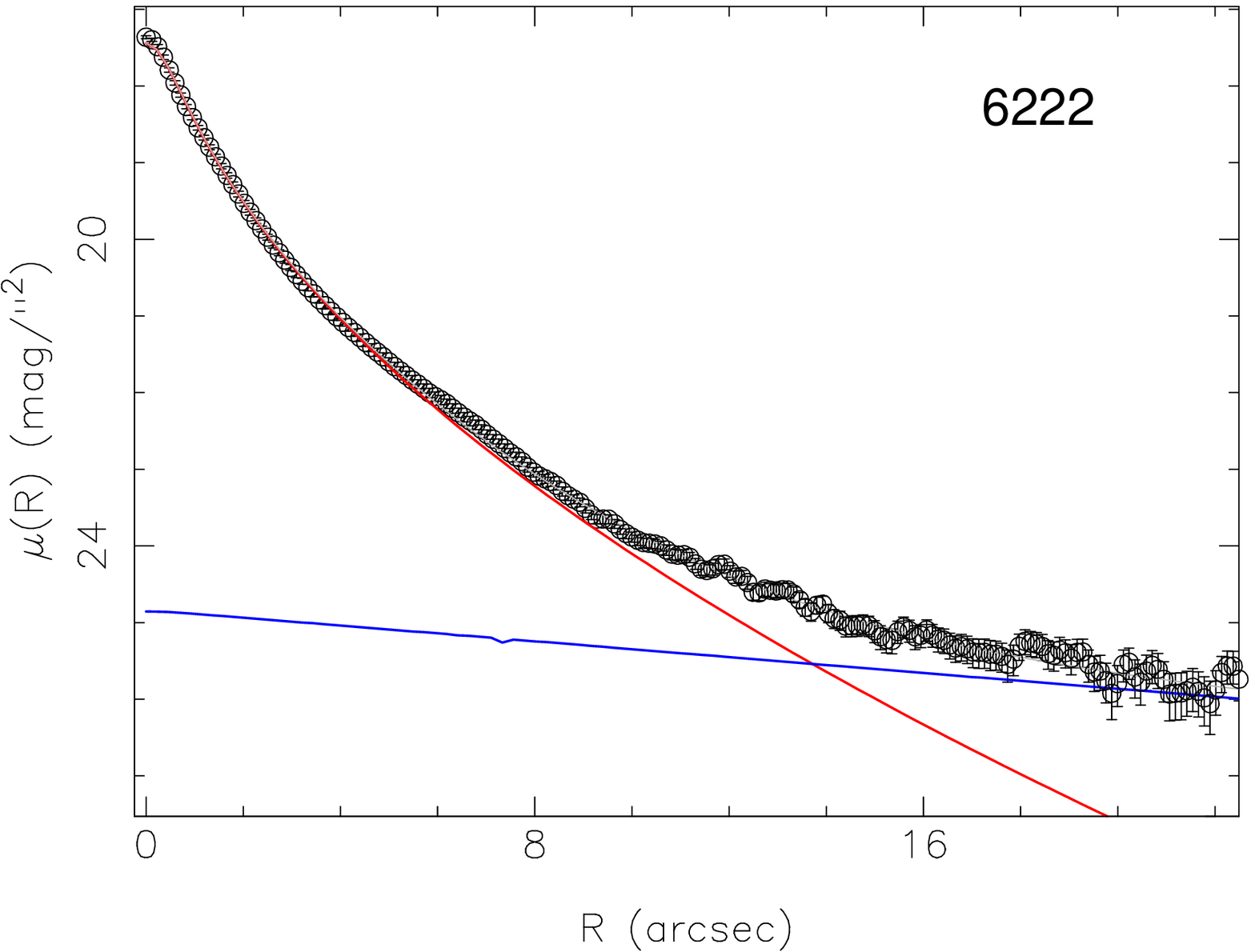}}
\centerline{\includegraphics[width=34mm]{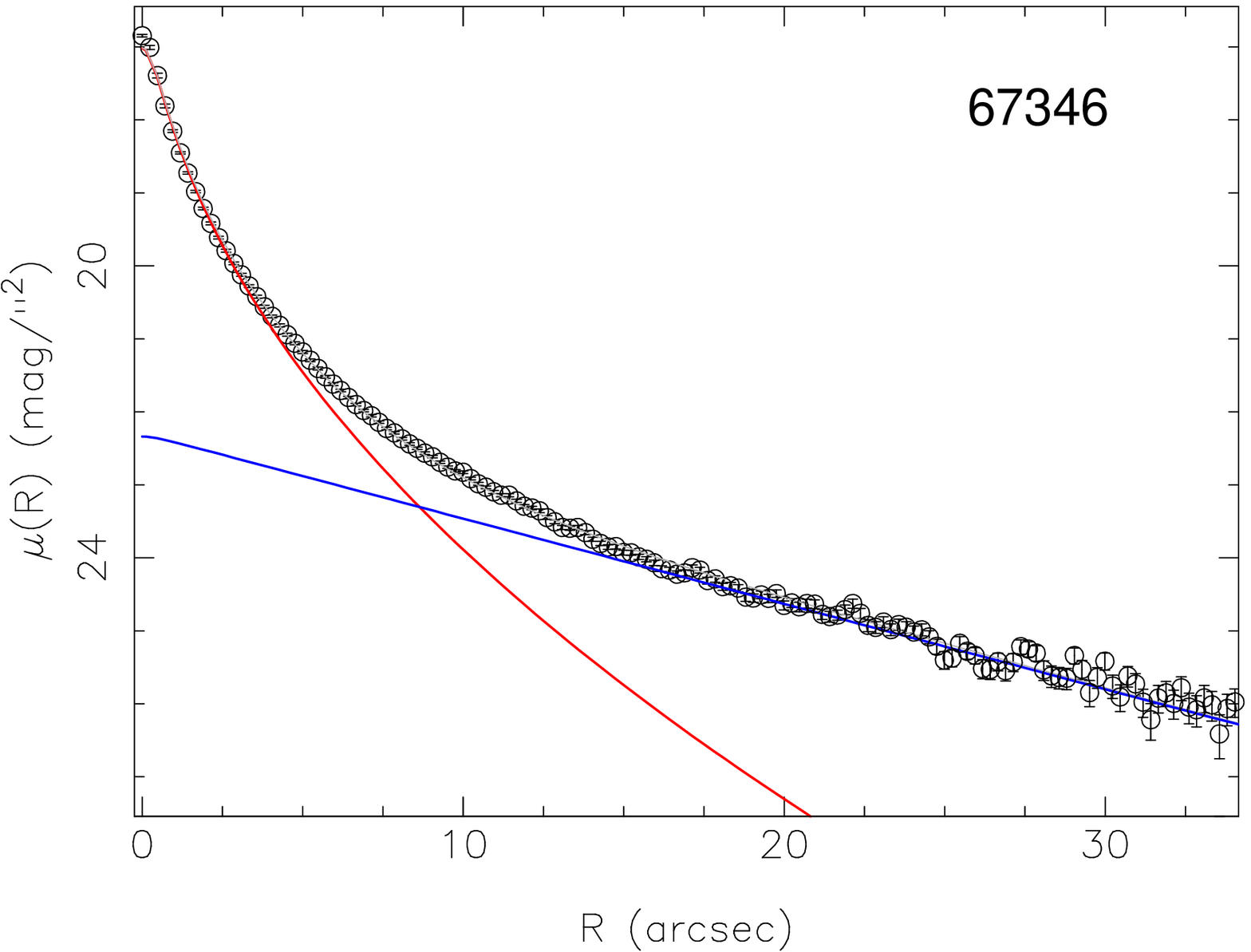}\includegraphics[width=34mm]{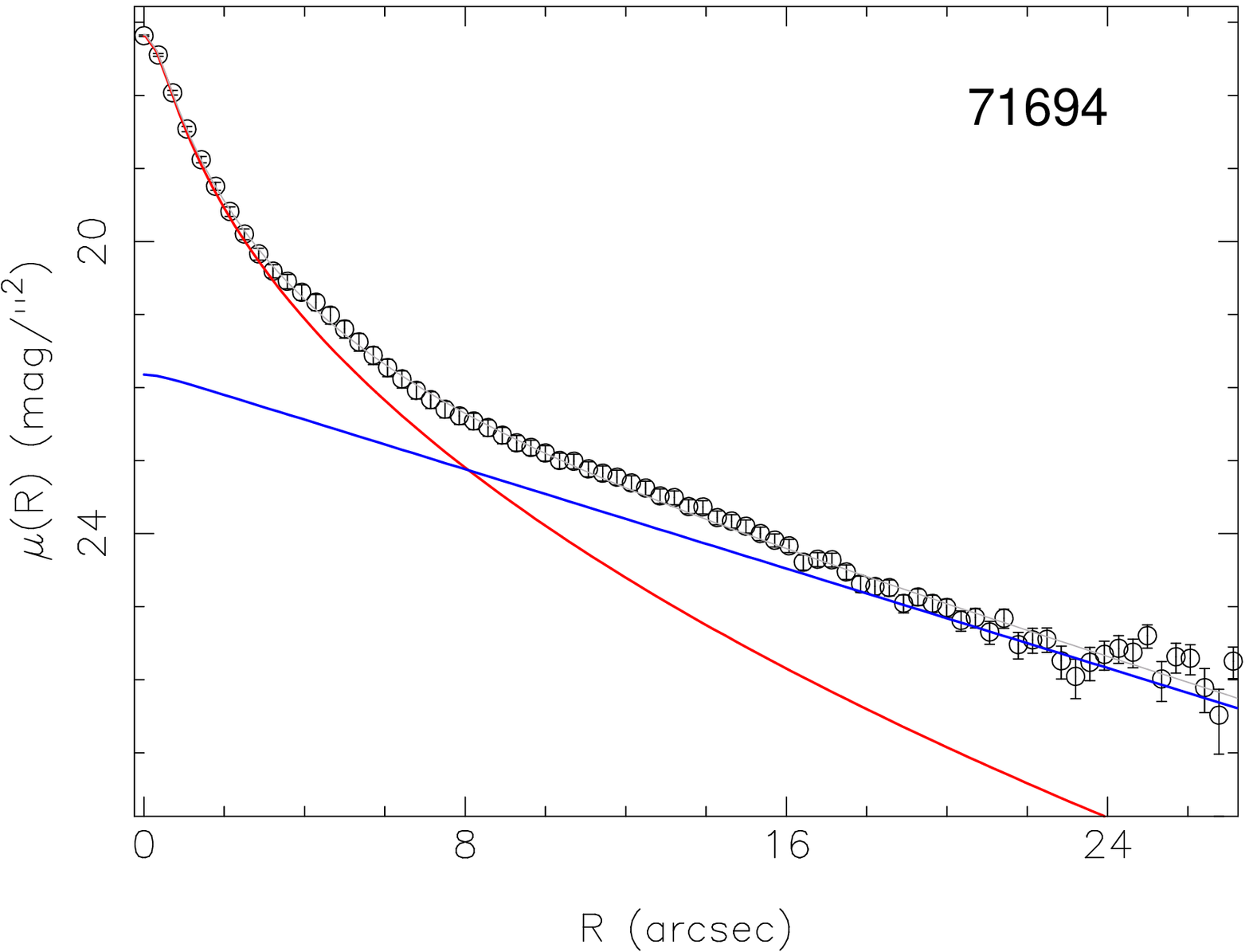}\includegraphics[width=34mm]{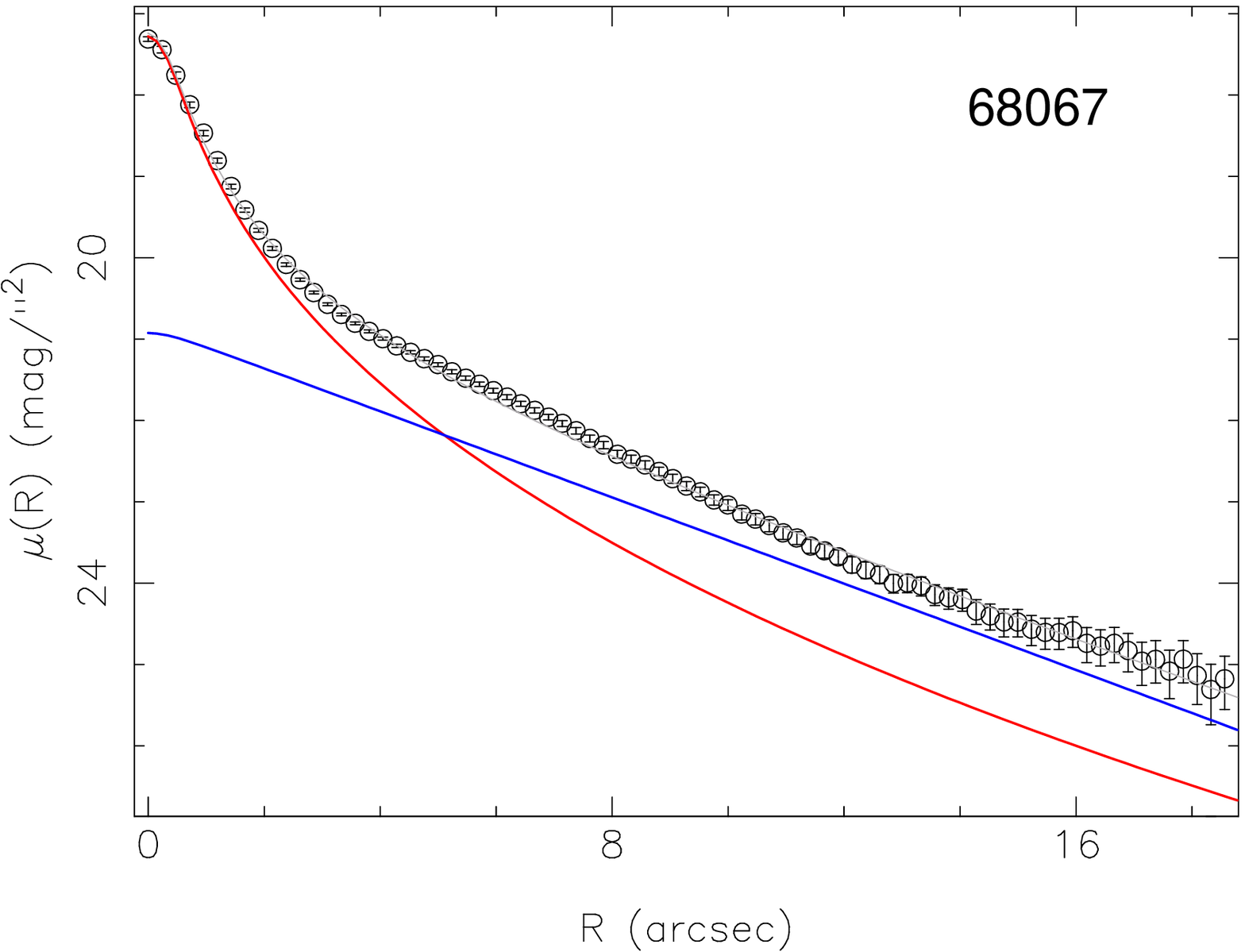}\includegraphics[width=34mm]{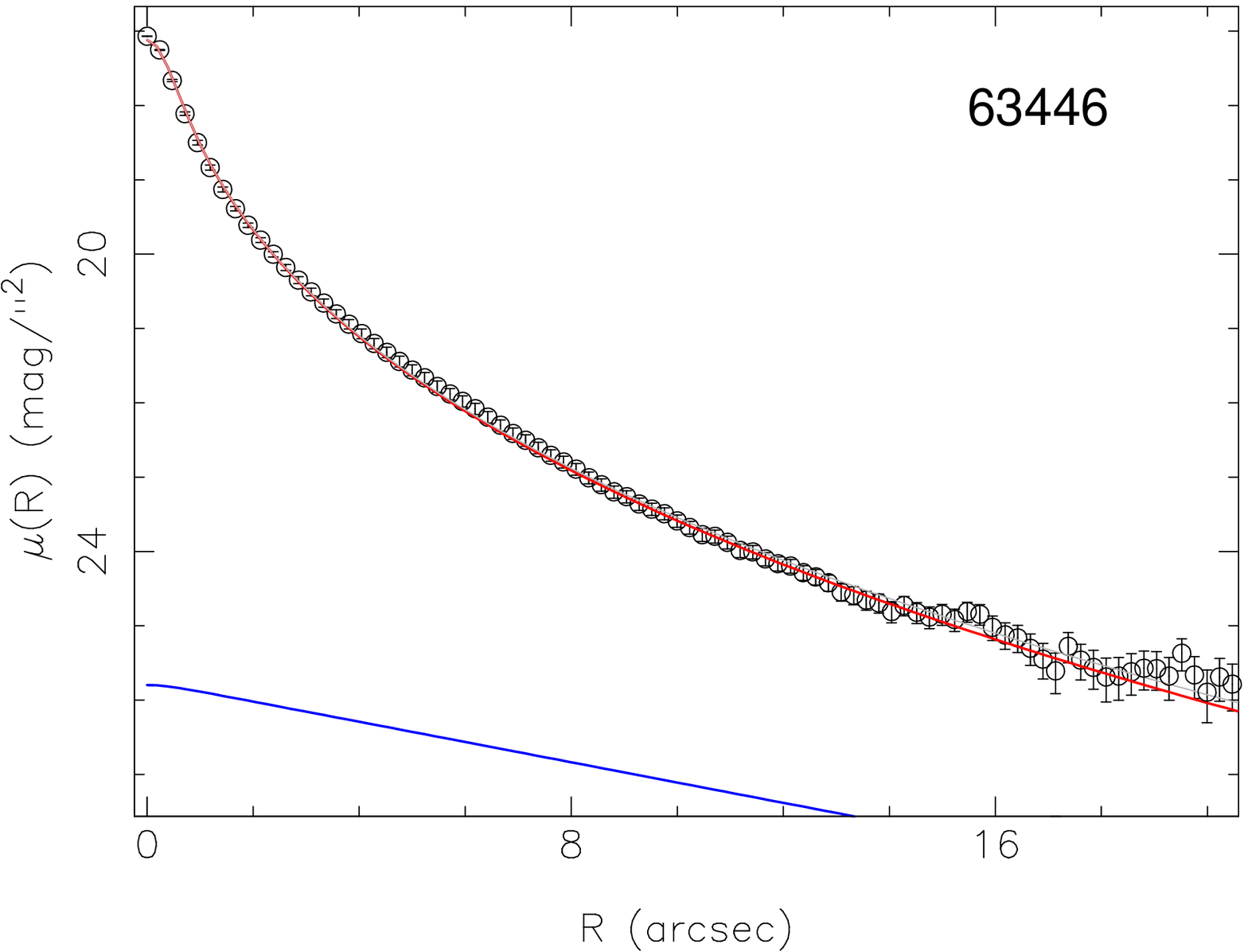}\includegraphics[width=34mm]{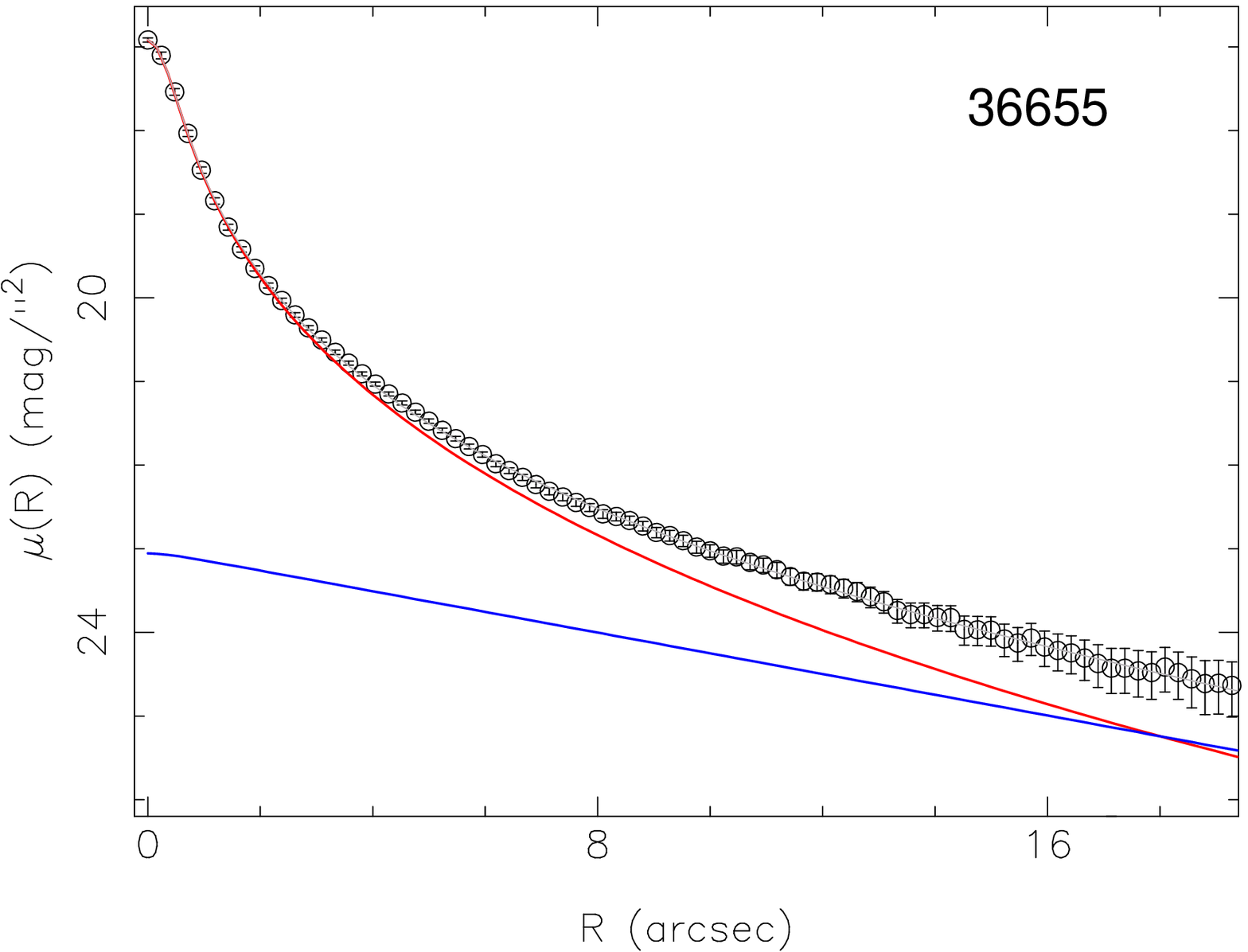}}
\caption{The bulge-disk decomposition of the 70$\mu$m-excess galaxies in order of decreasing $f_{24}/f_{K}$, showing the $R$-band surface brightness profile, with resulting best-fitting bulge (red curve; Sersic-fit) and disk (blue curve; exponential) components.} 
\label{1dprofile}
\end{figure*}

As well as providing a robust constraint for the level of star formation within H{\sc ii} regions, the mid-infrared emission constrains the intensity of the radiation heating the dust, via the compactness parameter, $\mathcal{C}$, which is of the form
\begin{equation}
\mathcal{C} \propto \frac{\langle L_{*}(t)\rangle}{\langle R(t)^{2}\rangle},
\end{equation}
where $L_{*}(t)$ and $R(t)$ describe the time evolution of the stellar luminosities and radii of the H{\sc ii} regions \citep{groves}. Hence $\mathcal{C}$ provides a measure of the density of radiation within H{\sc ii} regions available for heating the dust. As is shown in Fig.~5 of \citet{groves}, the primary effect of varying $\mathcal{C}$ is to change the temperature of the dust, and systematically shift the wavelength of the far-infrared dust emission. As the compactness parameter decreases from $\log\mathcal{C}{=}6.5$ to $\log\mathcal{C}{=}4.0$, the peak wavelength of the far-infrared emission moves to progressively longer wavelengths from 50$\mu$m to 130$\mu$m, i.e. the dust temperature drops from ${\sim}6$0K to ${\sim}2$0K. The constraint on the compactness parameter for our galaxies comes from their $f_{70}/f_{24}$ colours, which can only be modelled by an extremely low compactness parameter $\log\mathcal{C}{\sim}4.0$. Such a low $\mathcal{C}$ value produces dust temperatures and far-infrared SEDs similar to those produced by a radiation field of intensity 1--1$0{\times}$ the local interstellar radiation field (ISRF), similar to that modelled for NGC\,5866 \citep{draine07}. In contrast, star-formation within H{\sc ii} regions produces a more intense radiation field, heating the surrounding dust to higher temperatures, resulting in compactness parameters $\log\mathcal{C}{\ga}5.5$ \citep{dopita11}, inconsistent with our SED model fits. If we consider the predicted far-infrared colours of a galaxy with emission solely from the general ISRF, the maximal $f_{70}/f_{24}$ ratio would occur for a radiation field ${\sim}10{\times}$ the local ``solar neighbourhood'' ISRF \citep{compiegne}.

This view that the anomalous $f_{70}/f_{24}$ colours require very low values of the compactness parameter $\mathcal{C}$ is consistent with the model predictions of \citet{li10} who show that the characteristics of the dust emission as observed by the $f_{70}/f_{24}$ colour are strongly linked to the surface density of star formation within a star-forming region ($\Sigma$SFR/M$_{\odot}\,{\rm yr}^{-1}\,{\rm kpc}^{-2}$), independent of metallicity (their Fig. 8). In this case the 2--3$\times$ increase in $f_{70}/f_{24}$ required to produce our 70$\mu$m-excess galaxies from the normal star-forming population implies a reduction in $\Sigma$SFR of ${\sim}5$0--1$00{\times}$. Similarly, \citet{calzetti10} shows that the integrated $f_{70}/f_{24}$ colours of normal star-forming galaxies (LIRGs excluded) are tightly anti-correlated with the star formation surface density (their Fig. 17), while \citet{popescu} show that the mid-infrared colours mainly depend on the ratio between the diffuse component from the ISM (heated by both young and old stars) and the clumpy component from the H{\sc ii} regions (the F factor in their formulation). In this latter work, the largest $f_{70}/f_{24}$ ratios are produced by a diffuse component heated only by evolved stars without a significant additional clumpy component. 

We note at this point that there are clear limitations on our ability to model and constrain the far-infrared emission of these galaxies. Firstly, given that we have just two data points (24$\mu$m and 70$\mu$m), we can only reasonably fit the far-infrared emission by a single-component model with one free parameter, $\mathcal{C}$. This means that we cannot consider more complex or realistic models which are able to fit separately the diffuse cool dust component related to the general interstellar medium and the hotter, clumpy component linked to star formation \citep[e.g.][]{popescu}. Moreover, the lack of coverage beyond 70$\mu$m means that we are not mapping the peak of the far-infrared emission from large dust grains, and so cannot constrain either the temperature or mass of the cool, diffuse dust component. In particular, we cannot distinguish between models in which the excess 70$\mu$m emission is due to an increase in the mass of the cold, diffuse dust component, or rather an increase in its temperature. 
While our mid-infrared observations are sufficient to show that the infrared (${\sim}1$0--1000$\mu$m) emission comes primarily from a cool diffuse component, further observations in at least one passband at ${\sim}2$00--500$\mu$m to map the Rayleigh-Jeans tail of the far-infrared emission are required to constrain the nature of this diffuse component, in a manner similar to that already achieved using {\em Herschel} PACS/SPIRE observations in five far-infrared bands over 70--500$\mu$m \citep[e.g.][]{kramer,rawle}.

\subsection{Morphologies}
\label{morph}

We morphologically classified each spectroscopic member of the SSC by eye, placing them onto the revised Hubble scheme \citep{devauc59,devauc63}, using the high-quality $B$- and $R$-band WFI imaging from the SOS (${\rm FWHM}{\sim}0.7^{\prime\prime}$), extending the earlier morphological classification performed by \citet{gargiulo} for the passively-evolving subsample. 
The angular resolution and depth \citep{sos1} of the data allows us to assign each galaxy into the following classes: E, E/S0, S0, S0/S and S. The spirals (S) are subsequently split into Sa, Sb, Sbc, Sc, Sg and Irregulars, where Sg are clearly spirals, but for which we could not make any more detailed classification. E galaxies have smooth radial profiles without discontinuities in surface brightness. S0 galaxies show a clear additional non-spheroidal (disk or lens) component, but no apparent spiral arms. 
E/S0 galaxies appear azimuthally symmetric, preventing us from visually distinguishing between Es and face-on S0s.
The above classification scheme is identical to that used by \citet{thomas} for a subset of 201 SSC members in the vicinity of A\,3558 and A\,3562 as part of the ESO Nearby Abell Cluster Survey (ENACS). This allows us to compare directly our morphologies with a completely independent set of classifications, albeit those of \citet{thomas} are based on poorer quality $B,R$-band imaging (1.2--2.0$^{\prime\prime}$) from the 1.54m Danish telescope at La Silla \citep{katgert}. In general we find excellent agreement between the two sets of classifications, although the improved image quality of our WFI images revealed a small fraction of galaxies classified as S0 or S0/S by \citet{thomas} to be spirals.

Of the 23 SSC galaxies with $(f_{70}/f_{24}){>}25$, two were classed as Es, four as E/S0s, 11 as S0s, two as Sa and four later-type galaxies. The four late-types are the four lowest luminosity galaxies with $K{>}14$. That the vast majority of these galaxies are of morphological class Sa or earlier would appear surprising given that all have 70$\mu$m detections, even if we are sampling an overdense region where early-type galaxies should dominate. If we consider the morphologies of all 165 70$\mu$m-detected SSC galaxies, 65 are E/S0s while 100 are spirals (Sa or later). Based on binomial statistics there is a probability of just $3.4{\times}10^{-4}$ of extracting 17 or more E/S0 galaxies out of 23 by chance from the global population. What is most notable is that for all these galaxies we see smooth surface profiles with no evidence of star-forming regions.  This is particularly surprising given their 70$\mu$m-based selection, as all previous morphological surveys of mid/far-infrared or radio-selected cluster galaxies have been completely dominated by late-types and mergers \citep{smail,balogh02,coia}.

To obtain a more quantitative measure of the morphological structures of these galaxies we have performed a bulge-disk decomposition using {\sc 2dphot} \citep{labarbera} to fit the $R$-band radial surface brightness profile of each galaxy. In Figure~\ref{1dprofile} we show the 1-d radial $R$-band surface brightness profiles of 10/11 galaxies (black points) from our sample of 70$\mu$m-excess galaxies with $f_{70}{>}30$mJy (we could not perform surface photometry on one galaxy due to it being near a bright star), in order of decreasing $f_{24}/f_{K}$ and the best-fitting linear combination of a Sersic profile (red curve; representing the bulge) and an exponential profile (blue curve; representing the disk), convolved with the local point spread function, as estimated from nearby non-saturated stars. In almost all cases we see the primary characteristics of S0s, with the central regions dominated by a high-surface brightness bulge component, which makes way to a dominant disk component characterised by a smooth exponential profile at larger radii. 

\begin{figure}
\centerline{\includegraphics[width=80mm]{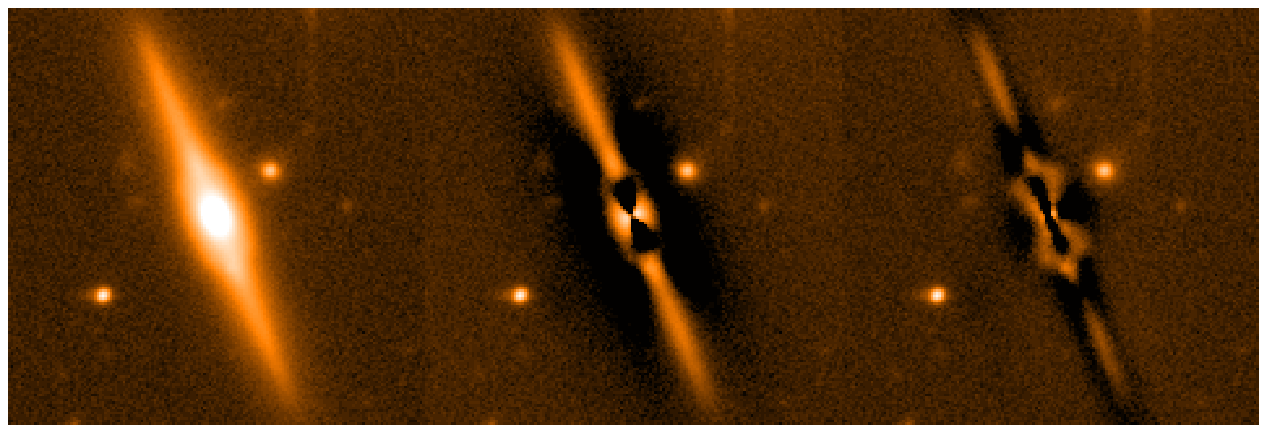}}
\centerline{\includegraphics[angle=270,width=80mm]{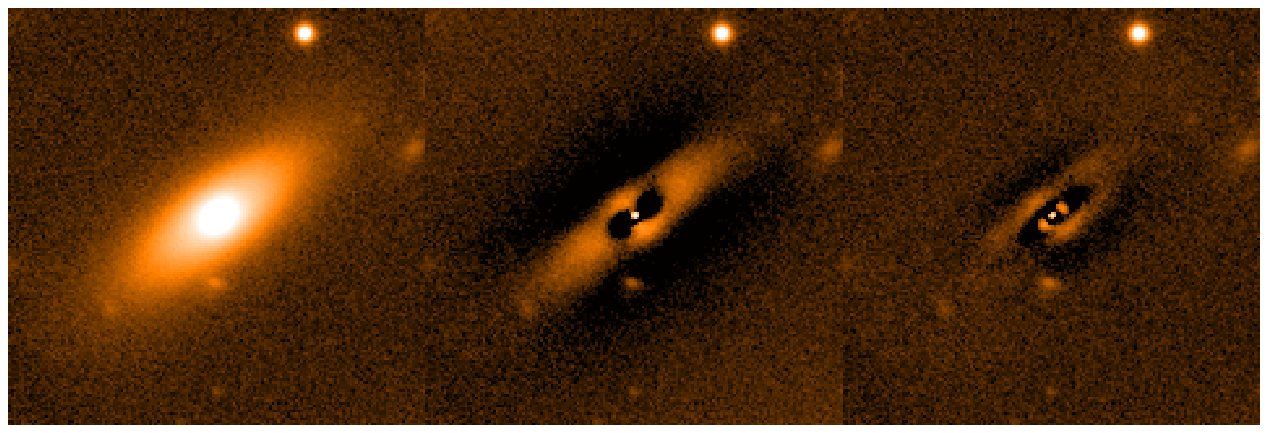}}
\centerline{\includegraphics[width=60mm]{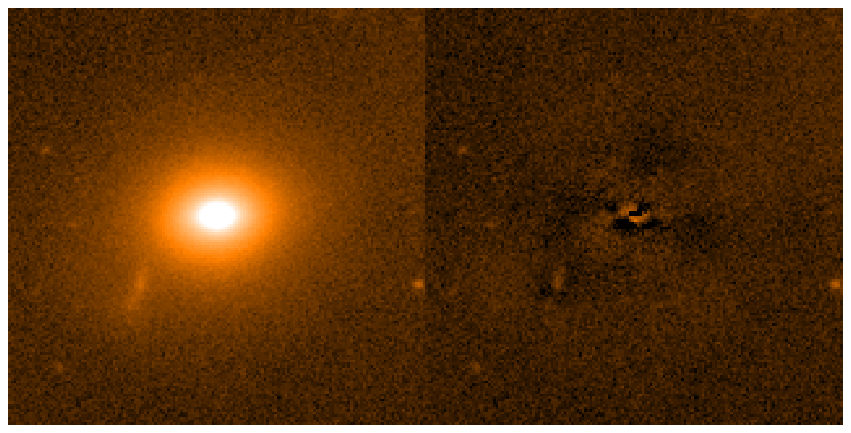}}
\centerline{\includegraphics[width=60mm]{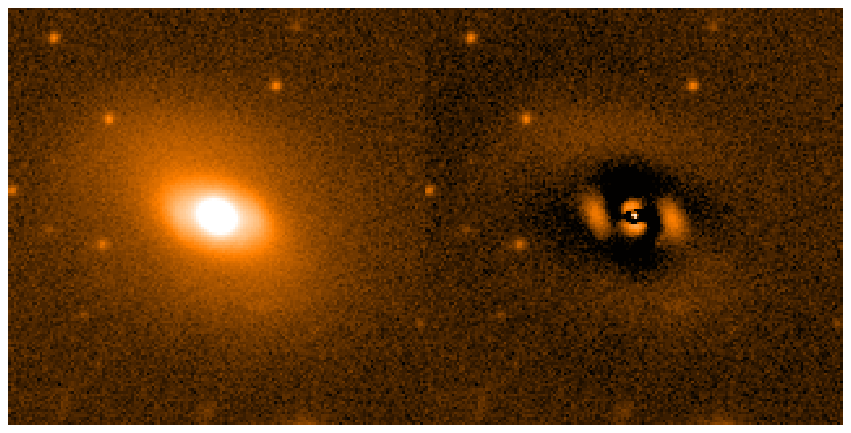}}
\caption{Sample two-dimensional fit results for 70$\mu$m-excess galaxies. Each plot shows the $R$-band galaxy thumbnail (left) and residual map(s) (centre/right) after subtraction of the best-fit GALFIT models), using the same grey-scale of intensity levels. The top two panels show the residual maps for both the best-fit single Sersic model (centre) and Sersic+exponential disk models, while the bottom two panels show the residual maps for just the Sersic+exponential disk model. The colour scaling is kept constant for each galaxy.}
\label{galfit}
\end{figure}

Looking at the bulge-disk decompositions in order of decreasing $f_{24}/f_K$ (from top-left to bottom-right) there visually appears some trend for the bulge component to become increasingly dominant while the disk component declines. Although within these 10 galaxies we only see a marginal correlation between the bulge-to-total (B/T) ratio and $f_{24}/f_{K}$, being significant only at an 88\% level based on the Spearman rank correlation test, when considering the whole 70$\mu$m-excess sample this significance increases to 97\%. This correlation suggests the star-formation is being quenched as the disk component declines and/or the bulge component is being built up in S0s. Alternatively, this trend could simply reflect that these are composite systems, comprising both bulge and disk components, with the 24$\mu$m emission associated to the star-forming disk.

We also performed full two-dimensional surface photometry of the 70$\mu$m-excess galaxies using the semi-automated {\sc Galfit} \citep[v2,][]{galfit} software. Thumbnails centred on each galaxy were extracted from the $R$-band images, and the surface photometry modelled as either exponential or Sersic profiles, or a linear combination of the two, each convolved with the local PSF, the best fit obtained by minimising the $\chi^{2}$ residuals. In Fig.~\ref{galfit} we show four example fits obtained by {\sc Galfit}. In the top two rows we show the original thumbnail image of the 70$\mu$m-excess galaxy (left panel), the residuals obtained after fitting with a single Sersic model (central panel), and the residuals obtained after fitting with a linear combination of a bulge and disk component (right panel). For these galaxies, we see that a single Sersic model doesn't provide an accurate fit to the galaxy profile, simultaneously underestimating the galaxy flux along the major axis, while over-subtracting the flux along the minor axis. The remnants of a disk component are clearly visible. The combination of bulge and disk components perform much better, certainly along the minor axis, although even here we see clear residuals from either a boxy bulge or spiral arms. In the lower two panels we simply show the original thumbnail image (left) and residual maps after subtracting the best-fit bulge+disk model. In the first case we obtain an excellent fit, with almost no residuals, while for the second case there is evidence for a remnant bar.

\subsection{The 70$\mu$m-excess phase along the quenching sequence.}

\begin{figure}
\centerline{\includegraphics[width=80mm]{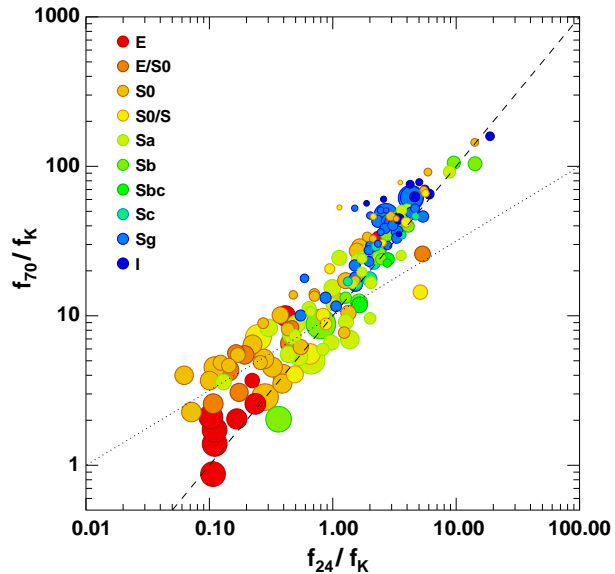}}
 \caption{The infrared $f_{70}/f_{K}$ versus $f_{24}/f_{K}$ colour-colour sequence for Shapley supercluster galaxies detected at 70$\mu$m. The symbols are coloured according to their morphological classification, while the sizes scale with the $K$-band luminosity. The dashed line indicates the sequence with $f_{70}/f_{24}{=}10$, while the dotted line indicates a sequence with $f_{70}/f_{K}{\propto}(f_{24}/f_{K})^{1/2}$}
\label{quenching}
\end{figure}

To place these 70$\mu$m-excess galaxies along the evolutionary pathway from normal star-forming spiral to passive S0s, we plot in Figure~\ref{quenching} $f_{24}/f_{K}$ versus $f_{70}/f_{K}$ flux ratios colour-coded according to their morphological classification. As discussed previously, we can consider the $f_{24}/f_{K}$ ratio to be a good proxy for the specific-SFR. The 70\,micron emission could be better described as a measure of the cooler dust (20--60K), either that belonging to the interstellar medium, or also that associated with the star-formation. For the spiral galaxy population (green/blue symbols) we can see the overall colour-colour trend can be well described a sequence in $f_{24}/f_{K}$ in which the 70\,micron emission is directly proportional to the 24\,micron emission ($f_{70}/f_{24}{=}10$; dashed line). This suggests that for these galaxies the 70\,micron emission produced primarily from dust heated by star-formation.

However, at the lowest $f_{24}/f_{K}$ ratios, we see the $f_{70}/f_{K}$ ratios of the S0s (orange/yellow symbols) lie systematically above this sequence (although ellipticals in contrast do not), suggesting that the 70\,micron emission is declining slower than the 24\,micron emission. 
Following \citet{doyon} who suggested that whereas the star-formation declines in a manner proportional to the H{\sc i} deficiency, the cool dust content declines at slower rate. If we consider the $f_{70}/f_{K}$ as a measure of the cool dust content, this would then decline as $(f_{24}/f_{K})^{1/2}$, as shown in Fig.~\ref{quenching} by the dotted line. As \citet{dacunha08,dacunha10} show, for actively star-forming galaxies stellar birth clouds are the main contributors to dust heating and hence the infrared emission, while in more quiescent galaxies, the bulk (${\ga}80$\%) of dust emission arises from the heating of dust by older stars in the interstellar medium. Indeed, for galaxies of a given stellar mass, they find $(M_{d}/{\rm SSFR}){\propto}{\rm SSFR}^{-1/2}$, and so for galaxies in which the 70\,micron emission and dust heating is primarily due to evolved stars, we may expect the dust content and hence $f_{70}/f_{K}$ to follow the dot-dashed curve. Interestingly, in a comparable {\em Spitzer}/MIPS analysis of 225 E/S0 galaxies, \citet{temi09} find a remarkably tight ``banana-shaped'' relation among the infrared colours formed by the $f_{24}/f_{K}$ and $f_{70}/f_{K}$ ratios. For E/S0s with $f_{24}/f_{K}$ ratios lying within the transition region up to the normal star-forming sequence, shows the $f_{70}/f_{K}$ ratio declining as $(f_{24}/f_{K})^{2/3}$ before turning over and declining rapidly as the galaxy reaches the passive sequence in $f_{24}/f_{K}$. 

The trends revealed by Fig.~\ref{quenching} support the 70$\mu$m-excess population representing a fundamental aspect of the quenching of star formation within early-type galaxies.

\section{Discussion}\label{sec:discuss}

We examined the mid-infrared colours of 165 70$\mu$m-detected spectroscopic members of the Shapley supercluster core at $z{=}0.048$, using panoramic {\em Spitzer}/MIPS 24$\mu$m and 70$\mu$m data. Among these galaxies we identified a population of 23 ``70$\mu$m-excess'' galaxies, comprising $14{\pm}3$ per cent of 70$\mu$m-detected SSC galaxies, whose mid-infrared colours $(f_{70}/f_{24}{>}25)$ are much redder than can be reproduced by any of the standard infrared SED models, and which also appear extremely rare in field surveys. These galaxies are primarily massive ${\sim}L^{*}_{K}$ galaxies with E/S0 morphologies, and intermediate $f_{24}/f_{K}$ ratios placing them in transition zone in which galaxies in the process of having their star formation quenched are likely to be located. 
In the two-component model for the far-infrared emission these colours could be understood as galaxies in which the star-formation is being quenched (or was recently), thus reducing the 24$\mu$m emission and explaining the low $f_{24}/f_{K}$ flux-ratios, while the emission seen at 70$\mu$m is largely due to dust heated predominately by relatively young stars ($10^{8}$--$10^{9}$\,yr) and the interstellar radiation field. 

These 70$\mu$m-excess galaxies comprise $8{\pm}2$ per cent of all $K{<}13$ ($M_{K}{<}M^{*}{+}1.3$) SSC members, indicating that they represent an important evolutionary pathway for cluster galaxies. Given this, we now attempt to place these galaxies within the theoretical frameworks provided by the various environmental processes that are currently believed to be important in transforming galaxies within clusters, including ram-pressure stripping and morphological quenching. 

\subsection{Ram-pressure stripping}

The concentration of these 70$\mu$m-excess galaxies in the cluster cores would suggest ram-pressure stripping as the mechanism causing the observed quenching of star-formation. In such galaxies, the gas is stripped from the outside-in, leaving truncated H{\sc i} disks \citep{vogt}. Within this truncation radius star formation continues normally, but outside evidence is found for recently quenched stellar populations \citep{crowl}. Using spatially-resolved {\em Herschel}/SPIRE maps \citet{cortese10} show the cold dust to be collocated with the gas, appearing truncated for highly H{\sc i}-deficient spirals in the Virgo cluster.
Hence, while star formation is quenched and gas removed via ram-pressure stripping, as the gas/dust surface densities within the truncation radius remain high, the star-formation {\em efficiency} should be unaffected, and the SFR and gas contents should decline in step. 

While the H{\sc i} gas contents of cluster galaxies may be partially or completely stripped by ram pressure, the molecular gas, by virtue of its much higher density and being more tightly bound within the potential well of the galaxy, should be much more difficult to remove. Indeed, in a CO survey 260 early-type galaxies from the volume-limited ATLAS$^{\rm 3D}$ sample, \citet{young} recently found no significant difference between the molecular gas contents of early-type galaxies in the Virgo cluster and those in isolated field regions within 24\,Mpc. Moreover, the dynamical status of the 12 CO-detected Virgo cluster galaxies (all S0s) is consistent with being a virialized population, deeply embedded within the hot intracluster gas, and unlikely to be recently accreted cluster members. They have probably retained their molecular gas over several Gyr within the cluster, albeit their molecular gas fractions (M$_{{\rm H}_{2}}/{\rm M}_{*}{\sim}0.$001--0.01) are now a factor ${\ga}10{\times}$ lower than those of normal star-forming galaxies \citep{saintonge}.

As ram-pressure stripping and the milder starvation mechanism prevent further accretion of pristine gas, the remaining H{\sc i} gas is enriched by metals recycled from stellar mass loss, particularly via the shedding of the stellar envelopes during the asymptotic giant branch phase, increasing the gas metallicity by a factor $\sim$2--3 \citep{skillman,boselli08}, which may in turn reduce the star formation efficiency (and hence 24$\mu$m emission) by a similar factor \citep{dib}. As the dust-to-gas ratio is known to scale approximately linearly with metallicity \citep{draine07}, we may expect the ratio between dust content and SFR to increase by a similar factor 2--3. This however requires us to make many assumptions about how the metals produce by evolved stars are frozen into dust grains as well as the rate at which dust grains are destroyed in the ISM for S0s passing through the cores of clusters as opposed to isolated field spirals. This enrichment of the remaining ISM by stellar mass loss could though be responsible for the anomalously red $f_{70}/f_{24}$ colours. Moreover an increase in the dust-to-gas ratio may directly reduce the star formation efficiency, if radiation pressure from the absorption and scattering of starlight by dust grains acts as an important mechanism in regulating star formation \citep{andrews}.

\citet{leitner} show that at low-redshifts gas recycling from stellar mass loss can provide most if not all of the gas required for star formation in galaxies, particularly those with lower than average specific-SFRs (as our galaxies are). The key for stellar mass loss to be efficient in fuelling star formation is the presence of at least some gaseous disk, into which the enriched material can be released directly. For elliptical galaxies with hot gas halos, the enriched gas released by AGB stars is likely rapidly disrupted and mixed with the hot halo gas. The simulations of  \citet{martig10} suggest that continuous recycling of gas through stellar mass loss also contributes significantly to the growth and survival of the disk component in these cluster S0s. \citet{davis} find that the molecular gas of the Virgo cluster CO-detected S0s is always kinematically aligned with the stellar kinematics, consistent with an internal origin such as stellar mass loss. 

\begin{figure}
\centerline{\includegraphics[width=80mm]{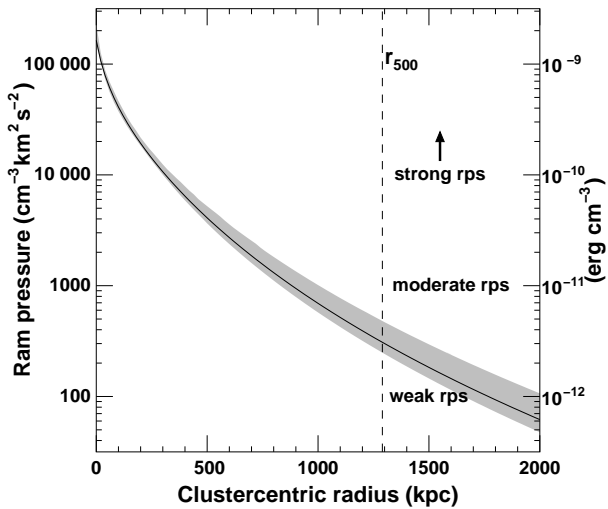}}
\caption{Effective ram pressure ($\rho_{ICM}\nu_{orb}^{2}$) acting on galaxies infalling into Abell 3558 as a function of cluster-centric radius (solid line), based on the gas/mass density profiles estimated by \citet{sanderson10} from {\em Chandra} X-ray data. The shaded regions indicate the 1$\sigma$ errors extrapolated from the uncertainties in the \citet{ascasibar} model parameters obtained from 200 bootstrap resamplings of the original X-ray data \citep[for details see][]{sanderson10}.}
\label{rps}
\end{figure}

Another explanation could be that the cold dust is heated during the stripping process, boosting the 70\,micron fluxes. This would be consistent with the 100$\mu$m-excess galaxies found by \citet{rawle} in the Bullet cluster, as the dust temperatures estimated by their fitting their combined {\em Herschel}/PACS+SPIRE photometry for these galaxies were found to be higher than expected given their infrared luminosity. The ram-pressure could also be removing the dust/gas from the vicinity of the source of heating (the stars). \citet{cortese10b} found extra-planar cold dust in the perturbed galaxy NGC\,4438 in Virgo collocated with the stripped atomic and molecular hydrogen. This extra-planar dust lacks a warm component or associated star formation, leading to unusual $f_{70}/f_{24}$ colours similar to those presented by our 70$\mu$m-excess galaxies. 

Finally, we check to see if ram-pressure stripping should be effective for these galaxies. For Abell 3558, \citet{sanderson10} derived the dark matter density, gas density ($\rho_{ICM}$) and gas temperature profiles by fitting \citet{ascasibar} models to the {\em Chandra} X-ray data. In these models, the dark matter is described by a \citet{hernquist} density profile, which provides simple analytic expressions for the mass profile, gravitational potential and escape velocity ($\nu_{esc}$). From this, we can derive the ram-pressure, $P_{ram}{=}\rho_{ICM}\nu_{orb}^2$ acting on galaxies as a function of cluster-centric radius, as shown in Fig.~\ref{rps}. For the orbital velocities, we assume the galaxies start at rest at the turnaround radius ($r_{ta}{\sim}3.5\,r_{200}$) and fall radially. 
We find obtain values of ${\sim}1\,000\,{\rm cm}^{-3}({\rm km\,s}^{-1})^{2}$ at 0.68\,$r_{500}$ \citep[where $r_{500}{=}1.29$\,Mpc;][]{sanderson}, indicative of moderate ram-pressure stripping, capable of partially stripping an $L^{*}$ spiral with rotational velocity $\nu_{rot}{\sim}1$50--200\,km\,s$^{-1}$ \citep[e.g.][]{vollmer,roediger,kronberger}, rising to ram pressures of ${\sim}10\,000\,{\rm cm}^{-3}({\rm km\,s}^{-1})^{2}$ at 0.24\,$r_{500}$, at which point all of the gas may be rapidly stripped from such galaxies. Thus for the bulk of our 70$\mu$m-excess galaxies, which lie within ${\sim}0$.5--1.0\,$r_{500}$, moderate-to-strong ram-pressure stripping should be effective. 

Furthermore, the velocity dispersion of the 13 70$\mu$m-excess galaxies within $r_{500}$ of Abell 3558 is 948\,km\,s$^{-1}$, marginally lower than the $975{\pm}39$\,km\,s$^{-1}$ obtained for the global cluster population within $r_{500}$. This would indicate that these 70$\mu$m-excess galaxies are a virialized population, having been embedded in the dense ICM for a Gyr or more, rather than being on their first infall into the cluster. The fact that they make up $8{\pm}2$ per cent of the cluster galaxy population would also imply that this 70$\mu$m-excess phase is long-lived, rather than the rapid evolution typically expected from ram-pressure stripping of infalling spirals, but as \citet{young} argue, the molecular gas content should be much more resistant to ram-pressure stripping than the easily removed H{\sc i} component. 

\subsection{Morphological quenching}

The 70$\mu$m-excess galaxies are morphologically identified as S0s, with smooth profiles and absent spiral arms. An alternative mechanism that could cause the decline in star-formation to occur more rapidly than the reduction in the dust/gas content could be morphological quenching. In this hypothesis, recently developed by \citet{martig}, the morphological transformation from spiral to lenticular makes the gas (and collocated dust) more stable against fragmentation and collapse into molecular clouds, leading to a quenching of star formation for a constant gas/dust fraction. The 24$\mu$m flux tracing the star-formation continues to decline, while the 70$\mu$m flux comes predominately from the cold interstellar dust content. The general interstellar radiation field in the bulge is predicted to be sufficiently strong to heat this dust without requiring young stars due to the high central density of stars \citep{jura}. IRAS detected far-infrared emission from a large fraction of nearby early-type galaxies, showing that they contained diffuse cool interstellar matter \citep{knapp,goudfrooij,merluzzi98}. 
Recent {\em Herschel} observations of nearby galaxies by \citet{engelbracht} and \citet{bendo} show that the temperatures of the cool dust component in the central regions of early-type spirals (S0--Sb) are 20--50\% higher than in the disks, indicating that the dust is heated by the higher stellar density in the galactic bulge. Again this dust heating would be consistent with the SED fits of \citet{rawle} to their 100$\mu$m-excess cluster population, although only one of their five 100$\mu$m-excess galaxies with HST imaging appears bulge-dominated.

While early-type galaxies are classically defined as gas-poor passive systems, molecular gas, H{\sc i} and warm dust was detected in 30 per cent of the SAURON representative sample of E/S0 galaxies \citep{combes}. \citet{crocker} resolved the molecular gas for 12 SAURON galaxies into central disks or rings, coincident with the dust distribution (often in the form of tightly-wound spiral structures) and corrotating with the ionized gas. However, star formation is the ionisation source for just half of the sample, corresponding to those with the largest molecular gas reservoirs. Most of their E/S0s do not lie on the FIR--radio relation, having excess FIR flux, while the bolometric IR flux produces SFR estimates ${\sim}3$ times higher than those based on the 8$\mu$m or 24$\mu$m fluxes, due to additional heating from evolved stars in the bulge. Interestingly, one galaxy from their sample, NGC\,4526 an S0 belonging to the Virgo cluster, is also a 70$\mu$m-excess galaxy with $f_{70}/f_{24}{\sim}30$, supporting the hypothesis that these are a cluster population. Of the 225 mostly field E/S0 galaxies in the {\em Spitzer}/MIPS study of \citet{temi09} only NGC\,5866 and NGC\,4526 have $f_{70}/f_{24}{>}30$, again suggesting that this is a rare or short-lived phenomenon in isolated field E/S0s. The above results all reveal that a fraction of early-type galaxies have significant gas and dust contents, producing emission in the far-infrared, at levels above those predicted given their extremely low levels of star formation.

\begin{figure}
\centerline{\includegraphics[width=80mm]{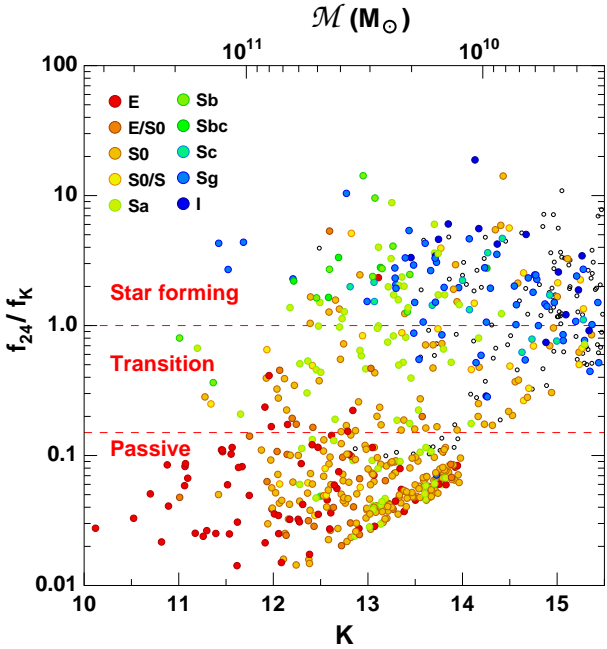}}
 \caption{The infrared colour $f_{24}/f_{K}$ versus $K$-band magnitude for confirmed Shapley supercluster galaxies, colour-coded according to their morphologiesas indicated. The red dashed lines indicate how we separate galaxies into star-forming, transition and passive populations according the their $f_{24}/f_{K}$ ratio as discussed in the text. Those galaxies not detected at 24$\mu$m are also plotted, after being given a random 24$\mu$m flux in the range 100--200$\mu$Jy, that is a factor 2--4 below the survey limit, and appear along the diagonal sequence passing through $K{\sim}13.5$, $f_{24}/f_{K}{\sim}0.05$.} 
\label{optmir_morph}
\end{figure}

\subsection{Caveats with both scenarios}

The presence of numerous S0s within the star-forming sequence of Fig.~\ref{optmir} and all along the quenching sequence of Fig.~\ref{quenching}, appears to point towards a morphological scenario in which spirals are first transformed into S0s without necessarily affecting their star-formation, before being quenched morphologically. This however, doesn't explain why these transition dusty S0s with litte or no star formation are located almost exclusively in the dense cluster cores and appear so rare in isolated field regions. In contrast the ram-pressure stripping scenario is able naturally to explain their concentration in the cluster cores where the dense ICM and high infall velocities make ram-pressure stripping most efficient. It doesn't however explain why these galaxies have already been morphologically transformed from spirals to S0s, as ram-pressure stripping is not predicted to strongly affect a galaxy's morphology. Instead, we would expect to find large numbers of spirals with little or no star-formation, which judging by Fig.~\ref{quenching} isn't the case. However we note that galaxies for which star-formation has been largely or completely quenched would not likely be detected at 70$\mu$m and hence aren't included in the figure.  

\subsection{Relating the quenching of star formation to the morphological transformation of galaxies within clusters}

To accurately account for the quenched galaxy population, and to be able to determine if indeed star-formation is primarily quenched {\em after} morphological transformation rather than before, we show in Fig.~\ref{optmir_morph} the infrared colour $f_{24}/f_{K}$ versus $K$-band magnitude of confirmed Shapley supercluster galaxies, colour-coded according to their morphological classifications, allowing us to directly relate the bimodality in the star-formation activity to morphology. As we would expect, the passive galaxies ($f_{24}/f_{K}{<}0.15$) are predominately early-types (E,E/S0,S0), while the normal star-forming galaxies ($f_{24}/f_{K}{>}1.0$) are mostly spirals (Sa or later). There are however notable populations of both passive spirals and star-forming S0s. 

Looking first at the 277 passive galaxies with $K{<}14$, we find that $10.1_{-1.5}^{+2.1}$ per cent are classified as spirals (27 Sa, 1 Sg). These examples of massive spirals with low SFRs are reminiscent of the classic ``anemic spiral'' of \citet{vandenbergh}, although all show also a prominent bulge. This would be consistent with \citet{bell} who show that truly quenched central galaxies (i.e. most massive galaxies within their host DM halo) have stellar bulges, indicating that a bulge, is an absolute requirement for full quenching of star formation (at least in central galaxies). For these passive spirals certainly ram-pressure stripping is a suitable candidate for removing their gas and quenching star-formation, in a manner similar to those spirals with truncated H$\alpha$-disks and/or H{\sc i} tails found in Virgo \citep{vogt,koopmann,crowl,chung}. 
Conversely, considering the 77 $K{<}14$ star-forming galaxies, $18.2_{-3.6}^{+5.2}$ per cent are classified as early types (2 S0/S, 10 S0, 1 E/S0, 1 E), for which it seems the morphological transformation has occured {\em prior} to any significant quenching of star-formation. This time sequence is the opposite of what has been suggested by previous studies \citep[e.g.][]{balogh00,bamford} which typically use optical colour as a proxy for star-formation activity. However, using {\em Spitzer} mid-infrared data \citet{wolf} revealed that these optically passive spirals still had significant obscured star formation, albeit at levels 0.6\,dex lower on average than normal star-forming galaxies of the same mass. They suggested that these apparent contraditions between the optical and mid-infrared viewpoints reflected that the unobscured star-formation in the outer disk is the most susceptible to quenching by ram-pressure stripping, while the obscured component in the central parts remains relatively unaffected. 

That there are comparable fractions and numbers of passive spirals and star-forming early-types indicates that star-formation in cluster galaxies is being quenched {\em both} before {\em and} after morphological transformation from spiral to S0. This is confirmed if we look at the morphologies of the transition galaxies with $0.15{<}(f_{24}/f_{K}){<}1.0$ likely to be in the process of having their star formation quenched. Here we find that $41.9_{-5.5}^{+5.9}$ per cent of the 74 $K{<}14$ transition galaxies are spirals (26 Sa, 2 Sb, 3 Sg), and $58.1_{-5.9}^{+5.5}$ per cent are early-types (29 S0, 2 S0-S, 6 E/S0, 6 E). Among this latter population, many have excess 70$\mu$m emission, producing mid-infrared colours inconsistent with those seen in the local field population or by model infrared SEDs. 

The concentration of these transitional S0s with excess 70$\mu$m emission towards the cluster cores points towards ram-pressure stripping being as the mechanism quenching the star formation in these galaxies. However, assuming the simple \citet{gunn} criterion the high central stellar densities of these S0s should significantly reduce the ability of ram-pressure stripping to completely remove their gas, while at low redshifts stellar mass loss should be capable of replenishing the gas supply requried to maintain observed low levels of star formation. As the findings of \citet{young} show, a significant fraction of S0s are able to retain their molecular gas for several Gyr within the dense ICM of the Virgo cluster. The apparent ``virialized'' dynamical status of the 70$\mu$m-excess population, suggests that these have also resided within the ICM for some time, rather than observed on their first infall. If this is the case, then it seems plausible that these are the direct descendants of the dusty starburst and post-starburst ${\sim}L^{*}$ galaxies abundant in clusters at $z{\sim}0.4$ \citep{poggianti99,dressler09}, but which appear rare in present day clusters.

\section{Summary}

We have presented an analysis of the mid-infrared colours of 165 70$\mu$m-detected spectroscopic members of the Shapley supercluster core at $z{=}0.048$, using panoramic {\em Spitzer}/MIPS 24$\mu$m and 70$\mu$m data. While the bulk of galaxies show mid-infrared colours typical of local star-forming galaxies, we identified a significant population of 23 ``70$\mu$m-excess'' galaxies, comprising $14{\pm}3$ per cent of 70$\mu$m-detected SSC galaxies, whose mid-infrared colours $(f_{70}/f_{24}{>}25)$ are much cooler than can be reproduced by any of the standard infrared SED models. We examined in detail the nature of these galaxies, finding the following:
\begin{itemize}
\item They are strongly concentrated towards the cores (${\la}0.5\,r_{500}$) of the five clusters that make up the SSC, and also appear extremely rare among local field galaxies from the SINGS and SWIRE surveys, confirming them as a cluster-specific phenomenon. 
 
\item Their intermediate $f_{24}/f_{K}$ ratios place them in transition zone in which galaxies in the process of having their star formation quenched are likely to be located. 
\item Their optical spectra and lack of significant ultraviolet emission imply little or no ongoing star formation.
\item They are primarily massive ${\sim}L^{*}_{K}$ galaxies with E/S0 morphologies, as confirmed by subsequent bulge-disk decompositions requiring comparable bulge and disk components to fit their $R$-band surface brightness profiles.
\item Fits to their panchromatic SEDs require their far-infrared emission to be produced mostly by the diffuse dust component heated by the general interstellar radiation field rather than the clumpy component related to ongoing star formation. Given that for most of these galaxies there is no evidence of ongoing or recent star formation, the origin of this dust heating mechanism remains unclear.
\end{itemize}

The joint analyses of the infrared colours and morphologies reveal that a primary evolutionary pathway for the formation of passive S0s in cluster cores involves the {\em prior} morphological transformation of spirals into S0s, perhaps via pre-processing or the impact of cluster tidal fields, before a subsequent quenching of star formation once the S0 encounters the dense environment of the cluster core. Many of the cluster S0s in this latter stage of transformation, that is in the process of having their star formation quenched, are characterised by excess 70$\mu$m emission, indicating that their cold dust contents are probably declining at a slower rate than their star formation. This observation, their concentration within the cluster cores, and their apparent virialized dynamical status, draw strong parallels with the population of molecular gas-rich S0s identified by \citet{young} in the Virgo cluster. These results suggest then that S0s are able to retain both their molecular gas and dust contents for a significant length of time (${\ga}1$\,Gyr) within the ICM, after being accreted by a cluster. These thus represent a key transition population that can be used to understand the physical mechanisms behind the formation of many passive lenticulars in local clusters. 

We put forward two possible causes for these cluster S0s with excess 70$\mu$m emission: (i) assuming the dust-to-gas ratio remains constant, the 70$\mu$m excess suggests a reduction in the star-formation efficiency in these dusty cluster S0s as proposed within the morphological quenching scenario; or (ii) a 2--3$\times$ increase in the dust-to-gas ratio or metallicity of the remnant ISM of these galaxies, as predicted by multizone chemical evolutionary models of galaxies in the process of being ram-pressure stripped or starved. In this latter scenario, the accretion of pristine gas is shut down, and so no longer dilutes the remaining H{\sc i} gas enriched by metals recycled by stellar mass loss. 

This evolutionary pathway is quite different from the often considered view that star formation is quenched in spiral galaxies once they encounter the cluster environment, forming the classic anemic spiral population \citep{vandenbergh,poggianti99,balogh00}, which are then transformed somehow into lenticulars. Notably, we identify significant populations of both star-forming S0s and passive spirals which reveal that star-formation in cluster galaxies is being quenched both {\em before} and {\em after} morphological transformation from spiral to lenticular, and that both pathways are of comparable relevance. This was confirmed when we looked at the transition galaxies likely to be in the process of having their star formation quenched, finding them to made up of 42 per cent spirals (mostly Sa) and 58 per cent early-type galaxies, most of which were S0s. 

These results confirm that the environmental processes behind the SFR--density and morphology--density relations are complex and are far from being fully understood. The availability of far-infared data is allowing us now to probe directly the interstellar medium via its dust content, which given that the environmental processes behind the quenching of star formation in cluster galaxies are likely to act directly on the ISM rather than the star forming regions themselves, will be crucial in fully understanding the origin of the SF--density relation.

\section*{Acknowledgements}

The authors thank the referee A. Dressler for his constructive report and positive feedback regarding this work. 
This work was carried out in the framework of the collaboration of the FP7-PEOPLE-IRSES-2008 project ACCESS. CPH, RJS and GPS acknowledge financial support from STFC.  GPS acknowledges support from the Royal Society. CPH, GB and PM thank Mike Dopita and the Australian National University for their hospitality.
We thank Alastair Sanderson for providing the gas/mass density profiles and uncertainties for Abell 3558. CPH, GPS thank Tim Rawle, Eiichi Egami, Maria Pereira, Alessando Boselli and Samuel Bossier for useful discussions. This work is based in part on observations made with the Spitzer Space Telescope, which is operated by the Jet Propulsion Laboratory, California Institute of Technology under a contract with NASA.

\label{lastpage}
\end{document}